\documentclass[12pt,a4paper]{article}
\input epsf
\usepackage{amsmath}
\usepackage{amssymb}
\usepackage{graphicx}

\newcommand{\pint}{\makebox[0pt][l]{\hspace{3.1pt}$-$}\int}

\newcommand{\atopfrac}[2]{\genfrac{}{}{0pt}{}{#1}{#2}}

\newcommand{\alg}[1]{\mathfrak{#1}}

\newcommand{\indups}[1]{_{\mathrm{\scriptscriptstyle #1}}}
\newcommand{\gym}{g\indups{YM}}

\newcommand{\Tr}{{\rm Tr \,}}

\newcommand{\cO}{{\cal O}}
\newcommand{\cN}{{\cal N}}
\newcommand{\fldZ}{\mathcal{Z}}
\newcommand{\fldD}{\mathcal{D}}
\newcommand{\phm}{\phantom{-}}

\begin{document}
\thispagestyle{empty}
\begin{flushright}
{\sc\footnotesize hep-th/0603157} \\
{\sc\footnotesize AEI-2005-165}\\
\end{flushright}
\vspace{1cm}
\setcounter{footnote}{0}
\begin{center}
{\Large{\bf Integrability and Transcendentality \par}}\vspace{20mm}
{\sc Burkhard Eden and Matthias Staudacher}
\\[7mm]
{\it Max-Planck-Institut f\"ur Gravitationsphysik\\
     Albert-Einstein-Institut \\
     Am M\"uhlenberg 1, D-14476 Potsdam, Germany}\\ [2mm]
{\tt beden@aei.mpg.de, matthias@aei.mpg.de}
\\[30mm]
{\sc Abstract}\\[2mm]
\end{center}

\noindent{
We derive the two-loop Bethe ansatz for the
$\alg{sl}(2)$ twist operator sector of $\cN=4$ gauge theory
directly from the field theory. We then analyze a recently
proposed perturbative asymptotic all-loop Bethe ansatz in the limit
of large spacetime spin at large but finite twist, and find a
novel {\it all-loop} scaling function. This function obeys the
Kotikov-Lipatov transcendentality principle and does not depend
on the twist. Under the assumption that one may extrapolate back
to leading twist, our result yields an all-loop prediction for the
large-spin anomalous dimensions of twist-two operators.
The latter also appears as an undetermined function in a recent
conjecture of Bern, Dixon and Smirnov for the all-loop structure
of the maximally helicity violating (MHV) $n$-point gluon amplitudes
of $\cN=4$ gauge theory. This potentially establishes a direct link
between the worldsheet and the spacetime S-matrix approach.
A further assumption for the validity of our prediction is that 
perturbative BMN (Berenstein-Maldacena-Nastase) 
scaling does not break down at four loops, or beyond.
We also discuss {\it how} the result gets modified {\it if} 
BMN scaling does break down.
Finally, we show that our result qualitatively agrees
at strong coupling with a prediction of string theory.}


\newpage

\setcounter{page}{1}


\section{Introduction and Main Results}
\label{sec:intro}

There is mounting evidence that planar $\cN =4$ gauge theory might
be ``{\it exactly solvable}''.
For example, it was recently proposed that
higher-loop maximally helicity violating (MHV) $n$-point gluon amplitudes
should be iteratively expressible through the (regulated)
one-loop amplitudes
\cite{AnaBerDixKos}, \cite{BerDixSmi}. Based on sophisticated two-loop
\cite{BerRozYan,AnaBerDixKos} and three-loop \cite{BerDixSmi}
computations, a conjecture for the {\it all-loop}
MHV $n$-point gluon amplitudes ${\cal M}_n$
in $4-2\,\epsilon$ dimensions was formulated in \cite{BerDixSmi}:
%
\begin{equation}\label{npoint}
{\cal M}_n=\exp \left[\sum_{\ell=1}^\infty\,a_\epsilon^\ell
\left(f^{(\ell)}(\epsilon)\,M_n^{(1)}(\ell\,\epsilon)+C^{(\ell)}+
E_n^{(\ell)}(\epsilon)\right)\right]\, .
\end{equation}
Here $E_n^{(\ell)}(\epsilon)$ vanishes as $\epsilon \rightarrow 0$,
$C^{(\ell)}$ are finite constants, and
$M_n^{(1)}(\ell\,\epsilon)$ is the $(\ell\,\epsilon)$-regulated
{\it one-loop} $n$-point amplitude.
At $\epsilon=0$ we have
$\underset{\epsilon \rightarrow 0}{\lim}\,a_\epsilon=g^2$
where $g^2$ is defined as
\begin{equation}\label{convention}
g^2 \, = \, \frac{\gym^2 \, N}{8 \, \pi^2} =
\frac{\lambda}{8 \, \pi^2}\, ,
\end{equation}
and $\lambda$ is the 't~Hooft coupling.
Finally, the $f^{(\ell)}(\epsilon)$ are generated in the
$\epsilon \rightarrow 0$ limit by the function
\begin{equation}\label{link}
f(g)=4\,\sum_{\ell=1}^\infty\,g^{2\,\ell}\,
f^{(\ell)}(0)\, .
\end{equation}
This function is in fact related to the large spin anomalous
dimension of so-called leading-twist operators in the gauge theory
\cite{ST}. Alternative names are ``soft'' anomalous dimension,
or ``cusp'' anomalous dimension. The simplest representatives
of $\cN=4$ twist operators are found in the $\alg{sl}(2)$ sector:
\begin{equation}\label{ops}
\Tr \big( \fldD^s \fldZ^L \big)+\ldots \, .
\end{equation}
Here $\fldD$ is a light-cone covariant derivative, $\fldZ$ is one
of the three complex scalars of the $\cN=4$ model, $s$ measures
the spacetime spin, and the twist $L$ is to equal to, in this sector,
one of the $\alg{so}(6)$ R-charges.
Leading twist is $L=2$.
The dots in \eqref{ops} indicate that the true quantum operators
are complicated mixtures of states, where the $s$ covariant derivatives
may act in all possible ways on the $L$ fields $\fldZ$.
Mixing with multi-trace operators is suppressed in the planar theory.
The function $f(g)$ is obtained by considering the large spin
$s \rightarrow \infty$ limit of the anomalous scaling dimension
of the quantum operators \eqref{ops}, which is expected to scale
logarithmically as
\begin{equation}\label{log}
\Delta \, = s \; + \; f(g)\; \log(s) \; + \; O(s^0)\, .
\end{equation}
We will call $f(g)$ of \eqref{link},\eqref{log} the
{\it scaling function}. Note that the scaling structure
in \eqref{log} is a highly non-trivial structural property of the exact
finite $s$ expression for $\Delta=\Delta(s,g)$.
Individual Feynman diagrams contributing at intermediate stages of the
perturbative calculation of $\Delta$ certainly contain higher ($k>1$) powers
$\log^k(s)$. The one-loop $\cO(g^2)$
contribution to $\Delta$ was first computed in
\cite{GWGP,DO} for all $s$, and indeed behaves as in \eqref{log}.
The $\cO(g^4)$ two-loop answer was found in \cite{KL,KLV}, and the
$\cO(g^6)$ three-loop one, inspired by a full-fledged computation in QCD
\cite{MVV}, in \cite{KLOV}. Again, the result indeed scales
as in \eqref{log}, and the state of the art up to now has been
\cite{KLOV}
\begin{equation}\label{klov}
f(g) =  4 \, g^2 \, - \frac{2}{3}\, \pi^2\, g^4 \, +
\frac{11}{45}\, \pi^4\, g^6 +\ldots \, .
\end{equation}
Fascinatingly, this agrees via \eqref{link} with the three-loop $n=4$
calculation of \eqref{npoint} by Bern, Dixon and Smirnov \cite{BerDixSmi}.

There is also mounting evidence that planar $\cN =4$ theory might
be {\it ``exactly integrable''}. This means that the {\it spectral problem},
i.e.~the spectrum of all possible scaling dimensions $\{\Delta\}$ of the
$\cN=4$ gauge theory,
is encoded in a {\it Bethe ansatz}.
This was established for the complete set of possible operators
of $\cN=4$ theory at one loop \cite{MZ,BS1}. It was then conjectured,
based on two- and three-loop computations in the model's
$\alg{su}(2)$ sector, that integrability extends to all orders
in perturbation theory, and, hopefully, to the non-perturbative
level \cite{BKS}. This was subsequently backed up in various studies
\cite{dynamic,SS,BDS,EJS,S,Z}, and culminated in a proposal for
the {\it asymptotic all-loop} Bethe equations of the theory \cite{BS2},
further supported by \cite{Sproofs}. Here ``asymptotic'' means
that the ansatz of \cite{BS2} is only expected to correctly yield the
anomalous dimension up to, roughly, $\cO(g^{2\,L-2})$, where $L$ is the
number of constituent fields (not counting covariant derivatives)
of the considered quantum operators\footnote{
In the case of twist $L=2$ operators the asymptotic ansatz actually
works to $\cO(g^6)$ instead of the naively expected $\cO(g^2)$
because of superconformal invariance \cite{S,BS2}.
}.
Very recently a proposal was made for circumventing the
asymptoticity restriction for the $\alg{su}(2)$ sector
by relating the dilatation operator, whose eigenvalues are the
dimensions $\Delta$, to a Hubbard Hamiltonian \cite{RSS}.
It would be exciting to also ``hubbardize'' the $\alg{sl}(2)$
sector of twist operators.

Let us stress that ``solvability'' and ``integrability''
are distinct concepts. The latter is a rather precise but narrow
concept referring to the spectral problem of the gauge theory.
It means that the spectrum is described by one-dimensional
factorized scattering of a set of appropriate elementary excitations.
Equivalently, it means that there is a Bethe ansatz. Exactly solving
the Bethe ansatz, in a given situation, is rarely easy, and
actually generically impossible. Integrability allows one to prove that
there will never be a ``plug-in'' formula for the spectrum of $\cN=4$ gauge
theory!

On the other hand, ``solvability'' is a significantly less precise concept.
Nevertheless, there is much evidence that in $\cN=4$ gauge theory
many quantities beyond the scaling dimensions allow for
a precise mathematical description. We have begun our discussion
with the conjecture \eqref{npoint}, which clearly contains
more than just spectral information. Another example are
coordinate space correlation functions of more than two local
composite operators. Certain intriguing iterative structures were
e.g.~noticed in four-point functions some time ago
\cite{4pt1,4pt2,4pt2Rome}. At the time of writing, the precise relation
between the observed solvable structures and the integrable structures
is somewhat reminiscent of the well-known
paradox of the chicken and the egg.

Recall that a three-loop $\alg{sl}(2)$ Bethe ansatz for gauge
theory was conjectured in \cite{S} 
by taking inspiration from the integrable structures 
appearing in string theory \cite{KZ},
and indeed reproduced \eqref{klov}.
We will further back up the conjecture of \cite{S} by calculating
in section \ref{factorized}
the two-loop S-matrix of the $\alg{sl}(2)$ sector directly from the
field theory. An alternative two-loop derivation, using
algebraic methods, was recently presented by Zwiebel \cite{Z}
(actually, for the bigger sector $\alg{su}(1,1|2)$).

It is amusing to note that \eqref{klov} is thus reproduced by
three completely independent procedures
(\cite{KLOV},\cite{S},\cite{BerDixSmi}),
none of them completely rigorous.
However, the various approaches, including their assumptions,
seem to be completely independent. So \eqref{klov} is very
likely to be correct!

In this paper we will apply the asymptotic all-loop
Bethe ansatz of \cite{BS2} in order to compute all further
perturbative corrections to the expression \eqref{klov}.
Strictly speaking, the asymptotic ansatz does not
apply to leading twist $L=2$, see above. We will however argue that
the scaling function \eqref{log} is universal in that it
describes the behavior of the lowest state of any $\alg{sl}(2)$
operator as long as $L \ll s$. For a very recent discussion of this
point, on the one-loop level, see \cite{korchemsky2}.
Our argument for the validity of our scaling function,
as concerns the leading twist operators, is therefore based
on {\it two} assumptions: (1) That it is indeed correct to
pick $L$ sufficiently large to stay in the ``asymptotic'' regime
of the Bethe ansatz, while keeping $L \ll s$, and (2) that
the Bethe ansatz of \cite{BS2} indeed describes the gauge theory,
for sufficiently ``long'' operators, at four loops and beyond.
The first assumption is very likely to be true, while the validity of the 
second is, at the time of writing, much less clear. However, our computation
might actually help to decide this issue, see below.

As a highly non-trivial check of our procedure, we will prove that
the anomalous dimension $\Delta$ is indeed of the expected scaling form
\eqref{log} to all order in perturbation theory.
We will find the scaling function to be given by the integral
representation
\begin{equation}\label{scalingfunction}
f(g) \, = \, 4 \, g^2 \, - \, 16 \, g^4 \, \int_0^\infty \,
dt\, \hat \sigma(t)\,
\frac{J_1(\sqrt{2}\, g\, t)}{\sqrt{2}\, g \, t} \, ,
\end{equation}
where the fluctuation density $\hat \sigma(t)$ is determined by the
solution of the integral equation
\begin{equation}\label{fredholm}
\hat \sigma(t) \, = \, \frac{t}{e^{\, t } - 1} \; \Bigl[ \;
\, \frac{J_1(\sqrt{2}\, g \, t)}{\sqrt{2}\, g \, t} \, -
2\, g^2 \int_0^\infty dt' \; \hat K(\sqrt{2}\, g \, t, \sqrt{2}\, g \, t') \;
\hat \sigma(t') \,
\Bigr]\, ,
\end{equation}
with the non-singular kernel
\begin{equation}\label{kernel}
\hat K(t, \, t') \, = \, \frac{J_1(t) \, J_0(t') \, - \, J_0(t) \, J_1(t')}
{t \, - \, t'}\, .
\end{equation}
The functions $J_0(t)$, $J_1(t)$ in the above equations are
standard Bessel functions.

We have been unable to find an explicit solution of the integral
equation. It would be quite interesting if this could be achieved.
It is however straighforward to obtain the weak-coupling expansion
of the fluctuation density $\hat \sigma(t)$ by iterating \eqref{fredholm}.
Using \eqref{scalingfunction} we then obtain the perturbative solution
of the scaling function $f(g)$. To e.g.~four-loop order one has
\begin{eqnarray}\label{prediction1}
f(g) & = & 4 \, g^2 \, - 4 \, \zeta(2) \, g^4 \, +  \,
\Bigl(4 \, \zeta(2)^2 \, + \, 12 \, \zeta(4)\Bigr) \, g^6 \\ && - \, \Bigl(4
\, \zeta(2)^3+ 24 \, \zeta(2) \zeta (4)-4 \, \zeta (3)^2+50 \, \zeta (6)
\Bigr) \, g^8  \, + \, \ldots\, , \nonumber
\end{eqnarray}
Using the fact that $\zeta$-functions of even argument may
be expressed as products of rational numbers and powers of $\pi$,
this may be simplified to
\begin{equation}\label{prediction2}
f(g) =  4 \, g^2 \, - \frac{2}{3}\, \pi^2\, g^4 \, +
\frac{11}{45}\, \pi^4\, g^6-
\left(\frac{73}{630}\,\pi^6-4\,\zeta(3)^2\right)\,g^8+\ldots \, .
\end{equation}

As a further non-trivial check of our procedure, and thus
the validity of \eqref{scalingfunction}, we shall find that
$f(g)$ obeys the Kotikov-Lipatov principle of maximal
transcendentality \cite{KL}, which was actually used in \cite{KLOV} in order
to extract, even at finite spin $s$, the $\cN=4$ dimensions from
the QCD calculation of \cite{MVV}. When applied to the
large $s$ limit, the principle holds that the sum over all the
arguments of the products of zeta functions appearing
as additive terms at a given loop order $\ell$ always adds up
to $2\,\ell-2$, see
\eqref{prediction1}, and \eqref{longprediction1} below.

It would be exciting if the four-loop prediction \eqref{prediction2}
could be tested by a field theoretic computation, maybe by way of
extending the results of \cite{BerDixSmi} to higher order.
Incidentally, even if field theory fails to reproduce \eqref{prediction2},
we will gain crucial knowledge on the integrable structure of the
gauge theory, see section \ref{breakdown}. The reason is that we are able
to predict {\it how} transcendentality will break down {\it if} it 
breaks down. The e.g.~four-loop term in the scaling function 
$f(g)$ in \eqref{prediction2} would then get modified to
\begin{equation}\label{prediction3}
- \left( \frac{73}{630}\,\pi^6-4\,\zeta(3)^2 + 8 \, \beta \, \zeta(3)
\right)\,g^8 \, ,
\end{equation}
where $\beta$ is an a priori unknown number\footnote{The alert reader
will notice that transcendentality could still be preserved if 
$\beta$ turned out to be a rational number times $\zeta(3)$.
The important point is that our calculation leads to a
{\it detection mechanism} for BMN-scaling violation.
If a future field theory calculation finds $\beta \neq 0$
BMN scaling breaks down.
}.
Furthermore this number would then
show up in the four-loop anomalous dimensions of {\it all} operators
of the $\cN=4$ theory, see \ref{breakdown}. 
In particular, it would manifest itself
in the four-loop dimensions of operators with a
large R-charge $J$, and would in fact induce a 
perturbative {\it breakdown} of BMN-scaling \cite{bmn}.
The argument may also be turned around: Proving that
\eqref{prediction2} holds as stated  
would establish that $\beta=0$, and would therefore be indirect
proof that BMN-scaling holds up to the four-loop level.

Finally, there is a prediction from string theory \cite{GKP2,FT},
assuming the AdS/CFT correspondence, for the {\it strong coupling}
$g \rightarrow \infty$ behavior of the scaling function
$f(g)$:
\begin{equation}
f_{\rm{string}}(g)=2\,\sqrt{2}\,g-\frac{3}{\pi}\,\log(2)
+\cO(\frac{1}{g})\, , \label{fstring}
\end{equation}
where $2\,\sqrt{2}\,g=\frac{\sqrt{\lambda}}{\pi}$, c.f.~\eqref{convention}.
The leading $\cO(g)=\cO(\sqrt{\lambda})$ piece is obtained from
a classical string spinning with a large angular momentum $s$ on the AdS 
space \cite{GKP2}, while the $\cO(g^0)=\cO(\lambda^0)$ term is the first
quantum correction obtained in \cite{FT}.

On the other hand, performing the strong coupling
limit for our scaling function as defined from the integral
equation \eqref{fredholm} with
\eqref{scalingfunction},\eqref{kernel} (see Section
\ref{strongsec}) we do vindicate the $O(g)$ asymptotics predicted
from string theory: the leading contribution to $\sigma$ is of
order $1/g^2$ and eliminates the first term on the r.h.s.~of
\eqref{scalingfunction}. However, our analysis is currently not
precise enough to decide whether or not the subleading $O(g)$ term
matches \eqref{fstring}. We hope to present a more complete 
solution of the strong coupling problem in future work.

\section{The Factorized Two-Loop $\alg{sl}(2)$ S-matrix}
\label{factorized}

In this preliminary chapter we will recall the Bethe ansatz for
the $\alg{sl}(2)$ sector of $\cN=4$
{\it twist operators} \eqref{ops} at one loop \cite{BS1}
and beyond \cite{S,BS2}. We will then derive it at two loops
by Feynman diagram computations, successfully checking part of the
conjecture of \cite{S}. A complimentary two-loop approach
was recently accomplished by Zwiebel \cite{Z},
who worked out the full (even non-planar) dilatation operator
of the bigger $\alg{su}(1,1|2)$ sector by algebraic means, and
also demonstrated the emergence of the two-loop two-body S-matrix
of \cite{S}.

The Bethe ansatz is obtained through the diagonalization of an
integrable spin chain, whose Hamiltonian is equivalent to
the dilatation operator. For a general introduction into
this technology see \cite{thesis}.
The states of the spin chain are represented by
removing the trace from the gauge theory states.
With $s_1+s_2+\ldots +s_{L-1}+s_L=s$, one has
\begin{equation}\label{replace}
\makebox[0cm][l]{\huge \hspace{-0.05em}\raisebox{-0.1em}{$\times$}} \Tr \Big( (\fldD^{s_1} \fldZ) (\fldD^{s_2} \fldZ) \ldots
(\fldD^{s_{L-1}} \fldZ) (\fldD^{s_L} \fldZ)\Big)
\longrightarrow
|s_1,s_2,\ldots,s_{L-1},s_L\rangle\, ,
\end{equation}
corresponding to a chain of length $L$. The $s_i$ are the spins of
the chain, and can, if $s$ is sufficiently large, take on any
value due to the noncompact character of the $\alg{sl}(2)$ sector.
The Hamiltonian of the chain acts on this state space.
Anomalous dimensions $\Delta$ are then related to the energies $E(g)$
(i.e.~the eigenvalues of the Hamiltonian) through
\begin{equation}\label{Delta}
\Delta=L+s+g^2\,E(g)\, .
\end{equation}

Recall the one-loop Bethe ansatz for $\alg{sl}(2)$,
corresponding to a XXX$_{-\frac{1}{2}}$
nearest-neighbor spin chain where the subscript indicates
a non-compact spin $-\frac{1}{2}$ representation of $\alg{sl}(2)$:
\begin{equation}\label{1loopbethe}
\left(\frac{u_k+\frac{i}{2}}{u_k-\frac{i}{2}}\right)^L=
\prod_{\textstyle\atopfrac{j=1}{j\neq k}}^s\,
\frac{u_k-u_j-i}{u_k-u_j+i}\, ,
\qquad \qquad
k=1,\ldots,s\, .
\end{equation}
The cyclicity constraint and the one-loop energy $E_0:=E(0)$ are
\begin{equation}\label{1loopmomeng}
\prod_{k=1}^s\,\frac{u_k+\frac{i}{2}}{u_k-\frac{i}{2}}=1
\qquad {\rm and} \qquad
E_0=\sum_{k=1}^s\, \frac{1}{u_k^2+\frac{1}{4}}\, .
\end{equation}
For a pedagogical derivation of these expressions from the Hamiltonian,
using coordinate-space Bethe ansatz, see \cite{S}.
A rigorous proof, for any representation
of $\alg{sl}(2)$, may be found in \cite{faddeev}.

The conjectured asymptotic all-loop Bethe ansatz for $\alg{sl}(2)$
\cite{BS2}
is then obtained by ``deforming'' the spectral parameter $u$, where the
deformation parameter is the Yang-Mills coupling constant $g$:
\begin{equation}\label{deformation}
u\pm\frac{i}{2}=x^\pm+\frac{g^2}{2 x^\pm}\, .
\end{equation}
It reads:
\begin{equation}\label{allloopbethe}
\left(\frac{x^+_k}{x^-_k}\right)^L=
\prod_{\textstyle\atopfrac{j=1}{j\neq k}}^s
\frac{x_k^--x_j^+}{x_k^+-x_j^-}\,
\frac{1-g^2/2x_k^+x_j^-}{1-g^2/2x_k^-x_j^+}\, ,
\qquad \qquad
k=1,\ldots,s\, ,
\end{equation}
with the new cyclicity constraint and the asymptotic all-loop energy $E(g)$
being given by
\begin{equation}\label{allloopmomeng}
\prod_{k=1}^s\,\frac{x^+_{k}}{x^-_{k}}=1
\qquad {\rm and} \qquad
E(g)=\sum_{k=1}^s\, \left(\frac{i}{x^+_{k}}-\frac{i}{x^-_{k}}\right)\, .
\end{equation}
It generalizes a three-loop Bethe ansatz first proposed in \cite{S}.
Very recently, much additional support of the ansatz was obtained
in \cite{Sproofs}. It should be noted, however, that we still
cannot currently prove that the ansatz
\eqref{allloopbethe},\eqref{allloopmomeng} really diagonalizes
the {\it gauge theory} at four loops and beyond. The reason
is that we do not know how to fix possible ``dressing factors''
(see \cite{Sproofs} and references therein.)

Directly proving the higher-loop ansatz from the gauge field
theory is hard. For all loops it will surely require ideas
that go far beyond ``summing up Feynman diagrams''.
To illustrate the complexity we will nevertheless derive
the Bethe ansatz at two loops by traditional methods.
Actually, we will be able to find the S-matrix, and we will
succeed in checking two-loop factorization of the three-body
problem. This is, according to \cite{GM}, a strong test
for integrability. Completing the proof would require
to demonstrate the factorization of the $s$-body problem
for arbitrary $s$, which we have not attempted to do.

\subsection{One-Loop Bethe Ansatz and Three-Body Factorization}
\label{bethe0}

The $sl(2)$ sector contains composite operators built from only one complex
scalar field $Z$ of the  ${\cal N} = 4$ SYM set of fields and the Yang-Mills
covariant derivative $D_\mu$. The operators are taken to carry symmetric
traceless irreps of the Lorentz group. We may project all indices onto
the complex direction $z = (x_1 + i x_2) / \sqrt{2}$, which guarantees
symmetrization while the trace terms automatically vanish.

Single trace operators of this type have a natural description as spin chains:
each field $Z$ is interpreted as an empty site which may be occupied
by any number of derivatives $D_z$. The spin chain Hamiltonian
\begin{equation}
{\cal H}^{(0)} \, = \, \sum_{i=1}^L \, {\cal H}_i^{(0)}
\end{equation}
involves a nearest neighbour interaction ${\cal H}_i^{(0)}$ that cyclically
acts on all sites of the chain of length $L$. Alternatively, one may consider
the asymptotic case, i.e.~an open chain of infinite length. The Hamiltonian can
transfer derivatives and it is conveniently expressed by matrices
containing amplitudes for such processes.

The one-loop Hamiltonian was worked out in \cite{oneloop}: let us denote the
number of derivatives on two adjacent sites as $\{s_1, \, s_2\}$. Then
\begin{eqnarray}
{\cal H}_i^{(0)}(\{s_1, \, s_2\} \, \rightarrow \, \{s_1, \, s_2\}) & = &
h(s_1) \, + \, h(s_2) \, , \label{nikdil} \\
{\cal H}_i^{(0)}(\{s_1, \, s_2\} \, \rightarrow \, \{s_1 - d, \, s_2 + d\})
& = & - \frac{1}{|d|} \nonumber
\end{eqnarray}
where $h(k)$ is the $k$-th harmonic number. The matrix elements refer
to a basis in which
$\{s_1, \, s_2\}$ is divided by
$s_1! \, s_2!$ in order to account for the indistinguishability of the
derivatives at each site.

The \emph{Bethe ansatz} rests on the observation that the
derivatives $D_z$ behave like particles (or ``magnons'') whose
motion is governed by a discrete Schr\"odinger equation. Let us
assign a position $x_i$ and a momentum $p_i$ to each magnon. One
constructs a wave function for each magnon number $s$:
\begin{eqnarray}
s \, = \, 1: & \sum_{x_1} & \Psi^{(0)}(x_1) \; |x_1\rangle \, , \\
s \, = \, 2: & \sum_{x_1 \leq x_2} & \Psi^{(0)}(x_1, \, x_2) \; |x_1, \, x_2
\rangle \, , \nonumber \\
s \, = \, 3: & \sum_{x_1 \leq x_2 \leq x_3} & \Psi^{(0)}(x_1, \, x_2, \, x_3)
\; |x_1, \, x_2, \, x_3 \rangle \, \ldots \nonumber
\end{eqnarray}
Here $|x_1\rangle$ denotes a state with a magnon at position $x_1$
etc.. The Hamiltonian reshuffles these ``kets'' as it can shift
magnons. On the other hand, the kets form a complete set of states
whose mutual independence one may use to transform the
Schr\"odinger equation
\begin{equation}
{\cal H}^{(0)} \sum_{x_1 \leq x_2 \leq \ldots} \Psi^{(0)}
(x_1, \, x_2, \, \ldots)
\, |x_1, \, x_2, \, \ldots \rangle \, = \, E^{(0)} \sum_{x_1 \leq x_2 \leq
\ldots} \Psi^{(0)}(x_1, \, x_2, \, \ldots) \, |x_1, \, x_2, \, \ldots \rangle
\end{equation}
into a difference equation on the wave function $\Psi^{(0)}$. We find
e.g. for only one magnon
\begin{equation}
2 \, \Psi^{(0)}(x_1) - \Psi^{(0)}(x_1-1) - \Psi^{(0)}(x_1+1) \, = \,
E^{(0)} \, \Psi^{(0)}(x_1)
\, . \label{omdifeq}
\end{equation}
This can be solved by Fourier transform:
\begin{equation}
\label{plane}
\Psi^{(0)}(x_1) \, = \, e^{i p_1 x_1} \, , \qquad E^{(0)} \, = \, 4 \,
\sin^2 \bigl(\frac{p_1}{2} \bigr) \, .
\end{equation}
The one-magnon problem thus defines the dispersion law, i.e.~the
dependence $E(p)$ of the energy on the momentum of the particle.
It is an essential assumption of the Bethe ansatz that the dispersion law
for several magnons is simply a sum over the contributions of the individual
pseudo-particles given by (\ref{plane}).

For two magnons the arguments of the wave function $\Psi^{(0)}(x_1, \, x_2)$
should obey $x_1 \, \leq \, x_2$ in order to avoid over-counting. Since the
one-loop Hamiltonian is a two-site interaction, the plane wave solution remains
valid when the separation of the magnons is greater or equal two. The
corresponding difference equation looks in fact like two copies of
(\ref{omdifeq}):
\begin{eqnarray}
&& 2 \, \Psi^{(0)}(x_1, \, x_2) - \Psi^{(0)}(x_1-1, \, x_2) - \Psi^{(0)}
(x_1+1, \, x_2) \\ & + & 2 \, \Psi^{(0)}(x_1, \, x_2) - \Psi^{(0)}(x_1,
\, x_2-1) - \Psi^{(0)}(x_1, \, x_2+1) \nonumber \\ & & = \, g^2 M \,
(E^{(0)}(p_1) \, + \, E^{(0)}(p_2)) \, \Psi^{(0)}(x_1, \, x_2) \, . \nonumber
\end{eqnarray}
It is a special feature of the Hamiltonian (\ref{nikdil}) that
this equation remains valid when $x_1 \, = \, x_2 - 1$.
However, we do find a new equation when $x_1 \, = \, x_2$ \cite{S}:
\begin{eqnarray}
&& \frac{3}{2} \, \Psi^{(0)}(x_1, \, x_2) - \Psi^{(0)}(x_1-1, \, x_2) -
\frac{1}{2} \, \Psi^{(0)}(x_1-1, \, x_2-1) \label{newcon} \\ & + &
\frac{3}{2} \, \Psi^{(0)}(x_1, \, x_2) - \Psi^{(0)}(x_1, \, x_2+1) -
\frac{1}{2} \, \Psi^{(0)}(x_1+1, \, x_2+1) \nonumber \\ & & = \, g^2 M \,
(E^{(0)}(p_1) \, + \, E^{(0)}(p_2)) \, \Psi^{(0)}(x_1, \, x_2)\, . \nonumber
\end{eqnarray}
A simple plane wave does not obey this equation, but we can solve
by an ansatz of the form
\begin{equation}
\Psi^{(0)}(x_1, \, x_2) \, = \, e^{i p_1 x_1 \, + \, i p_2 x_2} \, + \,
S^{(0)}(p_2, \, p_1) \, e^{i p_2 x_1 \, + \, i p_1 x_2} \, .
\label{psitwo}
\end{equation}
The physical intuition behind the last formula is that the
particles may scatter by exchanging their momenta; the second
plane wave is related to this, whereby the factor $S^{(0)}$ is called
the \emph{scattering matrix}. It can be determined from
(\ref{newcon}):
\begin{equation}
S^{(0)}(p_2, \, p_1) \, = \, - \frac{e^{i p_1 + i p_2} - 2 e^{i p_1} +
1} {e^{i p_1 + i p_2} - 2 e^{i p_2} + 1} \label{smat}
\end{equation}
Note that the two plane waves in (\ref{psitwo}) (with straight and
flipped momenta, respectively) are independent as functions.
Equation (\ref{newcon}) therefore yields two conditions, although
they are equivalent in this case.

For three magnons one writes an ansatz involving a wave function
$\Psi^{(0)}(x_1, \, x_2, \, x_3)$ subject to $x_1 \, \leq \, x_2 \, \leq \,
x_3$ and proceeds to set up difference equations. As before, the magnons do not
feel each other when $x_1 \, + \, 1 \, < \, x_2 \, < \, x_3 \, - \, 1$.
One might expect special behaviour when $x_1 \, = \, x_2 \, - \, 1$ or
$x_3 \, =  \, x_2 \, + \, 1$, but due to the structure of the Hamiltonian this
actually does not yield any new conditions. Thus it remains to investigate the
cases
\begin{equation}
(i) \; \; x_1 = x_2 < x_3 \, , \qquad (ii) \; \; x_1 < x_2 = x_3 \, ,
\qquad (iii) \; \; x_1 = x_2 = x_3 \, .
\end{equation}
We write an ansatz which straightforwardly generalizes the two-magnon formula
(\ref{psitwo}):
\begin{eqnarray}
\Psi^{(0)}(x_1, \, x_2, \, x_3) \, = & \phantom{+ \, S^{(0)}_{123}} \;
e^{i p_1 x_1
\, + \, i p_2 x_2 \, + \, i p_3 x_3} & + \, S^{(0)}_{132} \; e^{i p_1 x_1 \, +
\, i p_3 x_2 \, + \, i p_2 x_3} \\
& + \, S^{(0)}_{213} \; e^{i p_2 x_1 \, + \, i p_1 x_2 \, + \,
i p_3 x_3} & + \, S^{(0)}_{231} \; e^{i p_2 x_1 \, + \, i p_3 x_2 \, + \,
i p_1 x_3} \nonumber \\
& + \, S^{(0)}_{312} \; e^{i p_3 x_1 \, + \, i p_1 x_2 \, + \,
i p_2 x_3} & + \, S^{(0)}_{321} \; e^{i p_3 x_1 \, + \, i p_2 x_2 \, + \,
i p_1 x_3} \nonumber
\end{eqnarray}
If $x_1 \, = \, x_2 \, < \, x_3$, the difference equation can be separated
into three independent pieces according to which momentum multiplies $x_3$ in
the exponentials. The case $x_3 \, = \, x_2 \, > \, x_1$ obviously allows for
a similar distinction w.r.t. $x_1$. Five of the resulting six equations are
independent so that one may solve:
\begin{eqnarray}
S^{(0)}_{132} & = & S^{(0)}(p_3, \, p_2) \, , \\
S^{(0)}_{213} & = & S^{(0)}(p_2, \, p_1) \, , \nonumber \\
S^{(0)}_{231} & = & S^{(0)}(p_2, \, p_1) \, S^{(0)}(p_3, \, p_1) \, ,
\nonumber \\
S^{(0)}_{312} & = & S^{(0)}(p_3, \, p_1) \, S^{(0)}(p_3, \, p_2) \, ,
\nonumber \\
S^{(0)}_{321} & = & S^{(0)}(p_2, \, p_1) \, S^{(0)}(p_3, \, p_1) \, S^{(0)}
(p_3, \, p_2)
\, . \nonumber
\end{eqnarray}
This solution persists when all three magnons coincide, which is again a
non-trivial consequence of the structure of the Hamiltonian. We see that the
scattering remains non-diffractive, 
i.e.~the momenta are unaltered while they may
be exchanged between the magnons. What is more, the three-particle $S$ matrices
factor into two-particle processes.

\subsection{Bethe Ansatz and Three-Body Factorization at Two Loops}
\label{3body}

The original Bethe ansatz described in the last section may be
generalized to higher orders in perturbation theory \cite{S}. To this
end one writes a perturbation expansion of all relevant quantities,
namely the Hamiltonian, the
ingoing wave and the $S$ matrix. The central topic of this section is to
derive the two-loop correction to the $S$ matrix in the $sl(2)$ sector
directly from the ${\cal N}=4$ field theory,
and to check three-body factorization to two loops.

In the appendices \ref{app:twopoint} and \ref{app:dila} we derive the
two-loop Hamiltonian for one, two, and three
magnons from a graph calculation using ${\cal N} = 2$ superfields \cite{HSS}
and the SSDR scheme (supersymmetric dimensional reduction) \cite{DimRed}. The
supergraph formalism is preferable because it minimizes the number of Feynman
integrals; for the present purpose the ${\cal N}=2$ formulation is superior
to ${\cal N}=1$ supergraphs. We end up with a manageable calculation involving
about twenty graphs. SSDR is the best suited regulator since it allows one to
treat superfields in a version of dimensional regularization.\footnote{At the
current loop order the scheme cannot lead to ambiguities arising from the
$\epsilon$ tensor.}

We attack the problem of calculating quantities with open indices by
tensor decomposition and employ the QCD package \emph{Mincer} \cite{mincer} to
evaluate the resulting scalar integrals. The package uses $4-2 \epsilon$
dimensional vectors so that we explicitly have to symmetrize and take out
trace terms. This makes the computer algebra very awkward so that we have
limited the scope of the present work to low magnon numbers. The method was
detailed in \cite{mejan} by one of the authors. We will heavily draw upon
this reference in the appendices.\footnote{The calculation of the
two-loop anomalous dimension of the twist two operators \cite{KLV} uses
similar techniques; it is more fully automatized but renounces on the use of
supergraphs.} Appendix \ref{app:twopoint} reviews the
renormalization of two-loop two-point functions in dimensional regularization.
In Appendix \ref{app:dila} we introduce operators $\tilde D_1, \, \tilde D_2$
which generate the singular part of the one- and two-loop two-point functions
and we show that the second anomalous dimensions are matrix elements of the
combination $\tilde D_2 \, - \, 1/2 \, \tilde D_1^2$, thus reproducing the
two-loop effective vertex given in \cite{BKS}, where the renormalization of
the dilatation operator in dimensional regularization was first discussed.
Finally, the $\tilde D_i$ are constructed from the supergraphs and the
two-loop Hamiltonian is worked out for one, two, and three magnons.

In this section we display the Hamiltonian as it arises from
$\tilde D_2 \, - \, 1/2 \, \tilde D_1^2$ alone. One can introduce into it
a number of gauge parameters which do not appear in the difference equations
defining the wave function and the $S$ matrix. This freedom is (more than)
sufficient to make the Hamiltonian hermitian and to make the sum of all
elements in each row or column disappear, as was the case for the one-loop
dilatation operator. In Appendix \ref{app:dila}
we also give another set of transfer rules which
includes the contribution of a term $-1/4 \, [ {\cal H}_{1 \, i}^{(0)}, \,
{\cal H}_{0 \, i}^{(0)} ]$,
which is needed when the dilatation operator is required to reproduce the
$O(g^2)$ re-mixing of the $sl(2)$ sector operators. This term cannot
be made hermitian by the aforementioned gauge transformations and thus from
the point of view of the Bethe ansatz it is maybe best omitted. It is
interesting to note however, that the commutator term does not change the $S$
matrix, while it seems to make redundant any wave function renormalization in
the Bethe picture.

The disconnected pieces of the two-loop combinatorics do not influence the
Hamiltonian. The connected two-loop graphs can stretch over three
adjacent sites. The basis elements below denote the number of covariant
derivatives at these three sites; we have explicitly indicated a factor
$1/(s_1! \, s_2! \, s_3!)$ with which they were rescaled.
\vspace{0.5cm}
\newline
\begin{minipage}{15cm}
\textbf{Spin 1}  \\
basis: $\{\{1,0,0\},\, \{0,1,0\}, \, \{0,0,1\}\}$
\begin{equation}
{\cal H}^{(2)}_i(1) \, = \,
\left(
                 \begin{array}{lll}
                  -\frac{3}{4} & \phm 1 & -\frac{1}{2} \\
                  \phm 1 & -\frac{3}{2} & \phm 1 \\
                  -\frac{1}{2} & \phm 1 & -\frac{3}{4}
                 \end{array}
                 \right)
\nonumber
\end{equation}
\end{minipage}
\newline
\begin{minipage}{15cm}
\textbf{Spin 2} \\
basis: $\{\frac{1}{2}\{2,0,0\},\, \{1,1,0\}, \, \{1,0,1\}, \, \frac{1}{2}
\{0,2,0\}, \, \{0,1,1\}, \, \frac{1}{2} \{0,0,2\}\}$
\begin{equation}
{\cal H}^{(2)}_i(2) \, = \,
   \left(
   \begin{array}{llllll}
    -\frac{15}{32} & \phm \frac{19}{16} & - \frac{1}{2} & \phm \frac{1}{2} &
-\frac{1}{4} & -\frac{1}{16} \\
    \phm \frac{13}{16} & -\frac{5}{2} & \phm 1 & \phm \frac{23}{16} &
-\frac{1}{4} & -\frac{1}{4} \\
    -\frac{1}{2} & \phm 1 & -\frac{3}{2} & \phm 0 & \phm 1 & -\frac{1}{2} \\
    \phm \frac{1}{2} & \phm \frac{17}{16} & \phm 0 & -\frac{63}{16} &
\phm \frac{17}{16} & \phm \frac{1}{2} \\
    -\frac{1}{4} & -\frac{1}{4} & \phm 1 & \phm \frac{23}{16} & -\frac{5}{2} &
      \phm \frac{13}{16} \\
    -\frac{1}{16} & -\frac{1}{4} & -\frac{1}{2} & \phm \frac{1}{2} &
\phm \frac{19}{16} & -\frac{15}{32}
   \end{array}
   \right)
\nonumber
\end{equation}
\end{minipage}
\newline
\begin{minipage}{15cm}
\textbf{Spin 3} \newline
basis:
$\{\frac{1}{6} \{3,0,0\}, \, \frac{1}{2} \{2,1,0\}, \, \frac{1}{2} \{2,0,1\},
\, \frac{1}{2} \{1,2,0\}, \, \{1,1,1\}, \frac{1}{2} \{1,0,2\},$ \newline
$~~~~~~~~~~~\frac{1}{6} \{0,3,0\}, \, \frac{1}{2} \{0,2,1\}, \,
\frac{1}{2} \{0,1,2\}, \, \frac{1}{6} \{0,0,3\}\}$
\begin{equation}
{\cal H}^{(2)}_i(3) \, = \,
   \left(
   \begin{array}{llllllllll}
    \phm \frac{85}{288} & \phm \frac{115}{144} & -\frac{1}{2} & \phm
\frac{43}{72} & -\frac{1}{4} & -\frac{1}{16} & \phm \frac{71}{216} &
-\frac{1}{6} & -\frac{1}{24} & -\frac{1}{54} \\
    \phm \frac{1}{48} & -\frac{209}{96} & \phm 1 & \phm \frac{29}{24} &
-\frac{1}{4} & -\frac{1}{4} & \phm \frac{19}{24} & -\frac{1}{6} &
-\frac{7}{48} & -\frac{1}{24} \\
  -\frac{1}{2} & \phm 1 & -\frac{39}{32} & \phm 0 & \phm \frac{19}{16} &
-\frac{1}{2} & \phm 0 & \phm \frac{1}{2} & -\frac{1}{4} & -\frac{1}{16} \\
  \phm \frac{3}{8} & \phm \frac{29}{24} & \phm 0 & -\frac{247}{48} & \phm
\frac{17}{16} & \phm \frac{1}{2} & \phm \frac{109}{48} & -\frac{1}{12} &
-\frac{1}{6} & -\frac{1}{6} \\
    -\frac{1}{4} & -\frac{1}{4} & \phm \frac{13}{16} & \phm \frac{23}{16} &
-\frac{7}{2} & \phm \frac{13}{16} & \phm 0 & \phm \frac{23}{16} &
-\frac{1}{4} & -\frac{1}{4} \\
    -\frac{1}{16} & -\frac{1}{4} & -\frac{1}{2} & \phm \frac{1}{2} &
\phm \frac{19}{16} & -\frac{39}{32} & \phm 0 & \phm 0 & \phm 1 &
-\frac{1}{2} \\
 \phm \frac{71}{216} & \phm \frac{41}{72} & \phm 0 & \phm \frac{215}{144} &
\phm 0 & \phm 0 &
      -\frac{971}{144} & \phm \frac{215}{144} & \phm \frac{41}{72} & \phm
\frac{71}{216} \\
    -\frac{1}{6} & -\frac{1}{6} & \phm \frac{1}{2} & -\frac{1}{12} &
\phm \frac{17}{16} & \phm 0 & \phm \frac{109}{48} & -\frac{247}{48} &
\phm \frac{29}{24} & \phm \frac{3}{8} \\
    -\frac{1}{24} & -\frac{7}{48} & -\frac{1}{4} & -\frac{1}{6} & -\frac{1}{4}
      & \phm 1 & \phm \frac{19}{24} & \phm \frac{29}{24} & -\frac{209}{96} &
\phm \frac{1}{48} \\
    -\frac{1}{54} & -\frac{1}{24} & -\frac{1}{16} & -\frac{1}{6} &
      -\frac{1}{4} & -\frac{1}{2} & \phm \frac{71}{216} & \phm \frac{43}{72} &
      \phm \frac{115}{144} & \phm \frac{85}{288}
   \end{array}
   \right)
\nonumber
\end{equation}
\end{minipage}
\vspace{0.5cm}
\newline
Let us now focus on the Bethe ansatz. The spin-chain Hamiltonian up to two
loops is
\begin{equation}
{\cal H} \, = \, \sum_i \, {\cal H}^{(0)}_i \, + \, g^2 \, {\cal H}^{(2)}_i \,
+ \ldots \, , \qquad g^2 \, = \,
\frac{g^2_{YM} N}{8 \pi^2}
\end{equation}
and it has energy eigenvalues $E \, = \, E^{(0)} \, + \, g^2 E^{(2)} \,
+ \, \ldots \, $. The wave functions of the form
\begin{equation}
\Psi(x_1, \, x_2, \, \ldots) \, = \, \Psi^{(0)}(x_1, \, x_2, \, \ldots) \, +
\, g^2 \, \Psi^{(2)}(x_1, \, x_2, \, \ldots) \, + \, \ldots
\end{equation}
are contracted on the kets $|x_1, \, x_2, \, \ldots \rangle$ defined
in Section (\ref{bethe0}).

For one magnon we may scale away $\Psi^{(2)}$. The Schr\"odinger equation
\begin{equation}
\sum_i \, {\cal H}^{(2)}_i(1) \, \sum_{x_1} \, \Psi^{(0)}(x_1) \,
|x_1 \rangle \, = \,
E^{(2)} \, \sum_{x_1} \, \Psi^{(0)}(x_1) \, |x_1 \rangle
\end{equation}
leads to the difference condition
\begin{eqnarray}
&& - \frac{1}{2} \, \Psi^{(0)}(x_1 - 2) \, + \, 2 \, \Psi^{(0)}(x_1 - 1) \,
- 3 \, \Psi^{(0)}(x_1) \, + \, 2 \, \Psi^{(0)}(x_1 + 1) \, - \frac{1}{2} \,
\Psi^{(0)}(x_1 - 2) \nonumber \\
 && = \, E^{(2)} \, \Psi^{(0)}(x_1)
\end{eqnarray}
which can again be solved by Fourier transform:
\begin{equation}
\Psi^{(0)}(x_1) \, = \, e^{i p_1 x_1} \, , \qquad E^{(2)} \, = \, - 8 \,
\sin^4(\frac{p_1}{2}) \, .
\end{equation}
Hence the solution of the two-loop one-magnon problem yields the
correction to the one-loop dispersion law \eqref{plane}
for the magnon energy
$E(p)=E^{(0)}(p)+g^2\,E^{(2)}(p)+\cO(g^4)$. It is identical to the one of
the $\alg{su}(2)$ \cite{SS} and $\alg{su}(1|1)$ \cite{S} sectors,
and consistent with the proposed all-loop dispersion law of \cite{BDS} :
\begin{equation}
E(p)  =
\frac{1}{g^2} \Bigl( \sqrt{1 \, +\,
8 \, g^2 \, \sin^2(\frac{p}{2})} \, - \, 1 \, \Bigr) \, .
\nonumber
\end{equation}

The lowest order of the two magnon problem was discussed in the last section.
The two-loop part of the Schr\"odinger equation reads:
\begin{eqnarray}
&& \sum_i \, {\cal H}_i^{(0)} \, \sum_{x_1 \leq x_2} \, \Psi^{(2)}(x_1, \, x_2) \, | x_1, \, x_2
\rangle \\
& + & \sum_i \, {\cal H}^{(2)}_i \, \sum_{x_1 \leq x_2} \, \Psi^{(0)}(x_1, \, x_2) \, | x_1, \,
x_2 \rangle \nonumber \\
& = & (E^{(0)}(p_1) \, + \, E^{(0)}(p_2)) \, \sum_{x_1 \leq x_2} \, \Psi^{(2)}(x_1, \, x_2)
\, | x_1, \, x_2 \rangle \nonumber \\
& + & (E^{(2)}(p_1) \, + \, E^{(2)}(p_2)) \, \sum_{x_1 \leq x_2} \, \Psi^{(0)}(x_1, \, x_2)
\, | x_1, \, x_2 \rangle \nonumber
\end{eqnarray}
The resulting difference equations are perhaps not particularly elucidating. We
will rather comment on how to solve the system: the interaction length of the
two-loop Hamiltonian ${\cal H}^{(2)}_i$ is three. The two magnons must therefore behave as
free particles when $x_1 \, < \, x_2 - 2$. Thanks to the special form of ${\cal H}^{(2)}_i$
the same difference equation still holds when $x_1 \, = \, x_2 - 2$. The cases
of interest are thus
\begin{equation}
(i) \; \; x_1 \, = \, x_2 - 1 \, , \qquad (ii) \; \; x_1 \, = \, x_2 \, ,
\end{equation}
which both lead to new equations. In order to satisfy both conditions we must
allow for a correction not only to the $S$ matrix but also to the ingoing
wave function. Let
\begin{equation}
\psi(p_1, \, p_2) \, = \, (1 \, + \, g^2 \, \delta_{x_1, \, x_2} \,
f(p_1, \, p_2) ) \, e^{i p_1 x_1 \, + \, i p_2 x_2} \, .
\end{equation}
The wave function renormalization (``fudge factor'') is
local.
We write the ansatz
\begin{equation}
\Psi(x_1, \, x_2) \, = \, \psi(p_1, \, p_2) \, + \, S(p_2, \, p_1) \,
\psi(p_2, \, p_1) \, , \qquad S \, = \, S^{(0)} \, + \, g^2 \, S^{(2)}
\end{equation}
whose expansion in the coupling constant defines $\Psi^{(0)}, \, \Psi^{(2)}$.

Case (i) gives a condition relating $f(p_1, \, p_2)$ to $f(p_2, \, p_1)$.
Substituting this into (ii) makes the fudge factors disappear from the equation
so that we can solve for the $S$ matrix:
\begin{equation}
S^{(2)}(p_2, \, p_1) \, = \,
   -\frac{8 i \, \sin(\frac{p_1}{2}) \, \sin(\frac{p_1-p_2}{2}) \,
\sin(\frac{p_2}{2}) \, \left(\sin^2(\frac{p_1}{2}) \, + \,
\sin^2(\frac{p_2}{2})\right)}{(\sin
    (\frac{p_1-p_2}{2}) \, + \, 2 i \, \sin(\frac{p_1}{2}) \,
\sin(\frac{p_2}{2}))^2} \label{smat2loop}
\end{equation}
This result nicely confirms the conjecture for the two-loop
S-matrix of the $\alg{sl}(2)$ sector in \cite{S}.

The wave function renormalization $f$ is not fully determined. It is tempting
to assume it to be symmetric under the exchange of
$p_1, \, p_2$ since the magnons
are indistinguishable. In this case we find
\begin{equation}
f(p_1, \, p_2) \, = \, \sin^2\bigl(\frac{p_1}{2}\bigr) \, + \,
\sin^2\bigl(\frac{p_2}{2}\bigr)
\, - \, \frac{1}{2} \, \sin^2\bigl(\frac{p_1 + p_2}{2}\bigr) \, .
\label{fudgetwo}
\end{equation}
The alternative choice for the two-loop Hamiltonian from
Appendix \ref{app:dila} yields
$f(p_1, \, p_2) \, = \, 0$ if $f$ is symmetric.

The discussion of the two-loop three magnon scattering combines elements of
the one-loop three magnon case with the two magnon situation described in the
last paragraph. We write for the ingoing wave
\begin{eqnarray}
\psi(p_1, \, p_2, \, p_3) & = & \bigl(1 \, + \, g^2 M \, ( \, \delta_{x_1, \,
x_2} \, l(p_1, \, p_2, \, p_3) \, + \, \delta_{x_2, \, x_3}
\, r(p_1, \, p_2, \, p_3) \\ & & \phantom{ bigl(1 \, + \, g^2 M \, (} \,
+ \delta_{x_1, \, x_2} \, \delta_{x_2, \, x_3} \, u(p_1, \, p_2, \, p_3)) \,
\bigr) \, e^{i p_1 x_1 \, + \, i p_2 x_2 \, + \, i p_3 x_3} \nonumber
\end{eqnarray}
and make the ansatz
\begin{eqnarray}
\Psi(x_1, \, x_2, \, x_3) \, = & \phantom{+ \, S_{123}} \, \psi(p_1, \, p_2,
\, p_3) & + \, S_{132} \, \psi(p_1, \, p_3, \, p_2) \\ & + \, S_{213} \,
\psi(p_2, \, p_1, \, p_3) & + \, S_{231} \, \psi(p_2, \, p_3, \, p_1)
\nonumber \\ & + \, S_{312} \, \psi(p_3, \, p_1, \, p_2) & + \, S_{321} \,
\psi(p_3, \, p_2, \, p_1) \, , \nonumber
\end{eqnarray}
\begin{equation}
S_{ijk} \, = \, S^{(0)}_{ijk} \, + \, g^2 \, S^{(2)}_{ijk} \, . \nonumber
\end{equation}
As might be expected by now, the free situation must arise when $x_1 + 2 \, <
\, x_2 \, < \, x_3 - 2$ but in fact nothing changes when $x_1 + 2 \, = \, x_2$
or $x_2 \, = \, x_3 - 2$. We thus have to discuss the cases
\begin{eqnarray}
& (i)  & x_1 + 1 \, = \, x_2 \, < \, x_3 - 1 \, , \\
& (ii) & \phantom{1 + \; \, } x_1 \, = \, x_2 \, < \, x_3 - 1 \, , \nonumber \\
& (iii) & x_1 + 1 \, < \, x_2 \, = \, x_3 - 1 \, , \nonumber \\
& (iv) & x_1 + 1 \, < \, x_2 \, = \, x_3 \, , \nonumber \\
& (v) & x_1 + 1 \, = \, x_2 \, = \, x_3 - 1 \, , \nonumber \\
& (vi) & \phantom{1 + \; \, } x_1 \, = \, x_2 \, = \, x_3 - 1 \, , \nonumber \\
& (vii) & x_1 + 1 \, = \, x_2 \, = \, x_3 \, , \nonumber \\
& (viii) & \phantom{1 + \; \, } x_1 \, = \, x_2 \, = \, x_3 \, . \nonumber
\end{eqnarray}
In the first four cases only one $x$ has disappeared whereby one may use the
functional independence of the various exponential factors to organize each
difference equation into three separate constraints. Cases (i) and (ii)
are equivalent to a two-magnon problem with positions
$x_1$ and $x_2$: one may solve (i) for
three conditions relating $l(p_1, \, p_2, \, p_3)$ to $l(p_2, \, p_1, \, p_3)$
etc. and then substitute the three equations into (ii). This eliminates the
left fudge factor $l$ from the equations. Likewise, we can use (iii) to
eliminate the right fudge factor $r$ from (iv). We are left with six equations
on the five
$S^{(2)}_{ijk}$ matrices. A unique solution exists:
\begin{eqnarray}
S^{(2)}_{132} & = & S^{(2)}(p_3, \, p_2) \, , \\
S^{(2)}_{213} & = & S^{(2)}(p_2, \, p_1) \, , \nonumber \\
S^{(2)}_{231} & = & S^{(0)}(p_2, \, p_1) \, S^{(2)}(p_3, \, p_1) \, + \,
S^{(2)}(p_2, \, p_1) \, S^{(0)}(p_3, \, p_1) \, , \nonumber \\
S^{(2)}_{312} & = & S^{(0)}(p_3, \, p_1) \, S^{(2)}(p_3, \, p_2) \, + \,
S^{(2)}(p_3, \, p_1) \, S^{(0)}(p_3, \, p_2) \, , \nonumber \\
S^{(2)}_{321} & = & S^{(0)}(p_2, \, p_1) \, S^{(0)}(p_3, \, p_1) \, S^{(2)}(p_3, \, p_2)
\, + \,  S^{(0)}(p_2, \, p_1) \, S^{(2)}(p_3, \, p_1) \, S^{(0)}(p_3, \, p_2) \, +
\nonumber \\ & &  S^{(2)}(p_2, \, p_1) \, S^{(0)}(p_3, \, p_1) \, S^{(0)}(p_3, \, p_2)
\nonumber \, .
\end{eqnarray}
In other words, the complete $S$ matrix $S \, = \, S^{(0)} \, + \, g^2 \, S^{(2)}$
factors into two-particle processes also at two loops.

Once knowing that ${\cal H}^{(2)}_i(3)$ reproduces a two magnon problem when only two
arguments coincide, it is natural to put $l(p_1, \, p_2, \, p_3) \, =
\, f(p_1, \, p_2) \, = \, r(p_3, \, p_1, \, p_2)$ and so on. With these
identifications the cases (i) and (iii) reduce to the condition on the
two-magnon fudge factor $f$ found earlier. Of our remaining cases (v) is
empty while the last three all lead to one and the same condition on the
ultra-local fudge factor $u$. There is not enough information at this loop
order to solve for $u$ --- again, one may speculate that it should be chosen so
as to make the ingoing wave symmetric when all three positions coincide. The
solution is then similar to (\ref{fudgetwo}) if the two-loop Hamiltonian is as
defined in this section, or it vanishes for the alternative choice of ${\cal H}^{(2)}_i$ from Appendix \ref{app:dila}.

In conclusion, our analysis confirms the possibility of extending the $sl(2)$
sector Bethe ansatz to the two-loop level. It proves the functional form of the
two-loop $S$ matrix conjectured in \cite{S}, and it shows that the
three-magnon $S$ matrix factors into two-particle blocks.


\section{The Asymptotic All-Loop Large Spin Limit}

\subsection{One-Loop Large Spin Limit}

Consider the one-loop Bethe equations
\eqref{1loopbethe},\eqref{1loopmomeng} in the large spin limit
$s \rightarrow \infty$.
This problem was solved in great detail in the context of Reggeized
gluon scattering for the very similar case of a noncompact $\alg{sl}(2)$
spin$=0$ representation, i.e.~for a XXX$_{0}$ Heisenberg magnet,
in \cite{korchemsky}. The changes required to treat our present case of
noncompact $\alg{sl}(2)$ spin$=-\frac{1}{2}$ are minor.
Here we will proceed in a slightly different fashion as compared
to \cite{korchemsky}, where methods involving the Baxter-$Q$
function are employed. The reason is that the higher loop generalization
of the Baxter function is not yet known. We will therefore directly work with
the one-loop Bethe equations \eqref{1loopbethe},\eqref{1loopmomeng},
which nicely turn into a (singular) integral equation in the large spin limit.
Our method will then be extended to the asymptotic all-loop equations
\eqref{allloopbethe},\eqref{allloopmomeng} in the next section.
Interestingly, the effective higher-loop integral equation will
turn out to be non-singular.

Much intuition may be gained from the fact that the twist $L=2$ case is,
at one loop,
explicitly solvable for arbitrary spin $s$, cf.~Appendix \ref{app:hahn}.
Studying this solution one finds that the Bethe roots are all real\footnote{
It may be shown that, in contrast to the $\alg{su}(2)$ spin=$\frac{1}{2}$
Heisenberg magnet, the roots of the $\alg{sl}(2)$ spin=-$\frac{1}{2}$ Bethe
equations are, for all $L$ and $s$, always real.
We thank V.~Kazakov and K.~Zarembo for a discussion of this point.}
and symmetrically distributed around zero. The root distribution density
has a peak at the origin (in particular, there is no gap
around zero) and the outermost roots grow linearly with the spin as
$\max \{ |u_k| \} \, \rightarrow \, s/2 $. We therefore introduce
rescaled variables $\bar u$, and a density $\bar \rho_0(\bar u)$
normalized to one:
\begin{equation}\label{1loopdensity}
\frac{u_k}{s} \rightarrow \bar u
\quad {\rm with} \quad
\bar \rho(\bar u)=\frac{1}{s}\,
\sum_{k=1}^s\,\delta_0(\bar u-\frac{u_k}{s})\,
\quad {\rm and~thus} \quad
\int_{-\bar b}^{\bar b}\, \bar \rho_0(\bar u)=1\, .
\end{equation}
We now take the usual logarithm of the Bethe equations
\eqref{1loopbethe} and multiply either side by $-i$:
\begin{equation}\label{1looplog}
-i\,L\,\log \left(\frac{u_k+\frac{i}{2}}{u_k-\frac{i}{2}}\right)=
2\,\pi\,n_k
-i \sum_{\textstyle\atopfrac{j=1}{j\neq k}}^s\,\log
\frac{u_k-u_j-i}{u_k-u_j+i}\, .
\end{equation}
The integers $n_k$ reflect the ambiguity in the branch of
the logarithm, and may be interpreted as (bosonic) quantum mode numbers.
In the case of twist $L=2$ there is only one state. Its root
distribution is real and symmetric under $u \leftrightarrow -u$.
All positive (negative) roots have mode number
$n=1$ ($n=-1$). In the case of higher twist $L>2$ there is more than
one state\footnote{
The reader might find it instructive to consult Table 2 of \cite{S},
where a complete list of the three-loop spectrum of the first few states
of the $\alg{sl}(2)$ sector may be found.}.
However, for the lowest state the root distribution is again
real symmetric with $n=$sgn$(u)$.
Since $s$ is assumed
large, and $u_k=\cO(s)$ for nearly all roots, we furthermore expand
\eqref{1looplog} in $1/u$:
\begin{equation}\label{1looploglarge}
\frac{L}{u_k}=2\,\pi\,n_k-
2\,\sum_{\textstyle\atopfrac{j=1}{j\neq k}}^s\,
\frac{1}{u_k-u_j}\, .
\end{equation}
In this large $s$ limit the rescaled Bethe
roots condense onto a smooth cut on the interval $[-\bar b,\bar b]$ on the real
$\bar u$-axis. We may therefore take a continuum limit of \eqref{1looploglarge}
which yields, using \eqref{1loopdensity},
\begin{equation}\label{1loopsingular}
0=2\,\pi\,
\epsilon(\bar u)-2\,\pint_{-\bar b}^{\bar b} d\bar u'\,
\frac{\bar \rho_0(\bar u')}{\bar u-\bar u'}\, ,
\end{equation}
where $\epsilon(\bar u)=$sgn$(\bar u)$. In particular, the dependence on $L$
in \eqref{1looploglarge} drops out: The lowest state leads to the same
large $s$ root distribution, and therefore energy, for
{\it arbitrary} finite twist $L$.

The singular integral equation \eqref{1loopsingular}
is easily solved by inverting the
finite Hilbert transform with standard methods. The solution for
the rescaled one-loop root density is then found to be
\begin{equation}\label{1loopdenssol}
\bar \rho_0(\bar u)=
\frac{1}{\pi}\,\log \frac{1+\sqrt{1-4\,\bar u^2}}{1-\sqrt{1-4\,\bar u^2}}=
\frac{2}{\pi}\,\text{arctanh}\left(\sqrt{1-4\,\bar u^2}\right)\, ,
\end{equation}
where we have set the interval boundary to $\bar b=\frac{1}{2}$,
as obtained from the density normalization condition.
The result \eqref{1loopdenssol} of our procedure agrees with the
Baxter-$Q$ approach of \cite{korchemsky}.

Our derivation is closely modeled after the discussion of
\cite{BFST}; in particular, we refer to appendix C of that article.
There the ``spinning strings'' solutions of
\eqref{1loopbethe},\eqref{1loopmomeng}, where {\it both}
$s$ {\it and} $L$ are large and of the same order of magnitude
$\cO(L)=\cO(s)$, were studied. The difference is that in this case
the l.h.s~of \eqref{1looploglarge} is not negligible. The ensuing
potential $L/u$ on the l.h.s.~of \eqref{1looploglarge} opens up a
gap $[-\bar a,\bar a]$ of the root distribution in the
vicinity of $\bar u=0$. The resulting density for the lowest state
therefore has compact support
on two cuts $[-\bar b,-\bar a]$ and $[\bar a,\bar b]$ and is
 expressible through an elliptic
integral of the third kind (see eq.~(C.8) in \cite{BFST}).
One easily checks that when $L\rightarrow 0$ the gap disappears,
i.e.~$\bar a\rightarrow0$, and the elliptic density,
after rescaling the roots in \cite{BFST} by
$\bar u \rightarrow \frac{s}{L} \bar u$
in order to adapt conventions,
simplifies to the expression \eqref{1loopdenssol},
with $\bar b \rightarrow \frac{1}{2}$.

However, the one-loop anomalous dimension as obtained in \cite{BFST} does
{\it not} reproduce the expected logarithmic scaling of \eqref{log}
upon taking the limit $s/L \rightarrow \infty$.
Instead, it behaves like $\sim \log^2(s)$, cf.~(E.1) of \cite{BFST}.
This is a classic order-of-limits problem. Assuming $s,L$ large with
$s/L$ finite, and subsequently taking $s/L \rightarrow \infty$
does not yield the same result as taking $s$ large while keeping $L$
either finite or, at least, $L \ll s$.
For a very recent, quite extensive discussion of this fact
see \cite{korchemsky2}. 
For a recent study of some of the fine-structure of the
spinning strings limit see \cite{GK}.

The correct result is obtained by a careful derivation of the expression
for the energy in the continuum limit $s \rightarrow \infty$.
From the right equation in \eqref{1loopmomeng} we find,
using \eqref{1loopdensity},
\begin{equation}\label{scalingeng}
E_0=\frac{1}{s}\,\int_{-\frac{1}{2}}^{\frac{1}{2}} d\bar u\,
\frac{\bar \rho_0(\bar u)}{\bar u^2+\frac{1}{4\,s^2}}\, .
\end{equation}
Therefore, as opposed to the limit of \cite{BFST}
(see the expression in (C.4)) it is
nonsensical to use the unregulated expectation value
$\int d\bar u \, \bar \rho_0(\bar u)/\bar u^2$ for the energy.
The correct expression
\eqref{scalingeng} is actually related to the resolvent $G(\bar u)$, which
is defined for arbitrary complex values of $\bar u$ barring the interval
$[-1/2,1/2]$ (this integral is {\it not} of principal part type) as
\begin{equation}\label{resolvent}
G(\bar u)=\int_{-\frac{1}{2}}^{\frac{1}{2}} d\bar u'\,
\frac{\bar \rho_0(\bar u')}{\bar u'-\bar u}\, ,
\end{equation}
through
\begin{equation}\label{1loopeng}
E_0=\frac{2}{i}\,G\left(\frac{i}{2 s}\right)\, .
\end{equation}
Note that this further distinguishes the large spin limit
from the ``spinning strings'' limit, where the resolvent
generates the full set commuting charges
\cite{AS}.
One then finds from \eqref{1loopdenssol} that
\begin{equation}\label{resolventsol}
G(\bar u)=i\,\log \frac{\sqrt{1-4\,\bar u^2}+1}{\sqrt{1-4\,\bar u^2}-1}\, .
\end{equation}
Using now \eqref{1loopeng} and taking $s \rightarrow \infty$
we find
\begin{equation}\label{1loopengscaling}
E_0=4\,\log (s) +\cO(s^0)\, ,
\end{equation}
which is the well-known correct result, as may also be checked
directly from the exact finite $s$ result $E=4\,h(s)$,
see \eqref{hahneng}.

\subsection{Asymptotic All-Loop Large Spin Limit}

Let us now generalize the analysis of the previous section
to the higher loop case. We would therefore like to compute
the corrections to the one-loop density \eqref{1loopdenssol}
and energy \eqref{1loopengscaling} as generated by the deformed
Bethe equations \eqref{allloopbethe},\eqref{allloopmomeng}.
Compelling arguments for its validity to three loops were
presented in \cite{S} (in particular the equations reproduce
the conjecture of \cite{KLOV} based on the QCD calculation
\cite{MVV}, and they agree with \cite{BerDixSmi}).
Their all-loop form was conjectured in \cite{BS2}.
See also \cite{Sproofs}.

We begin by rewriting the asymptotic all-loop Bethe equations
\eqref{allloopbethe} with the help of \eqref{deformation} in the
following fashion:
\begin{equation}\label{allloopbethe2}
\left(\frac{u_k+\frac{i}{2}}{u_k-\frac{i}{2}}\right)^L\,
\left(\frac{1+g^2/2(x_k^-)^2}{1+g^2/2(x_k^+)^2}\right)^L
=
\prod_{\textstyle\atopfrac{j=1}{j\neq k}}^s
\frac{u_k-u_j-i}{u_k-u_j+i}\,
\left(\frac{1-g^2/2x_k^+x_j^-}{1-g^2/2x_k^-x_j^+}\right)^2\, ,
\quad
k=1,\ldots,s\, .
\end{equation}
Let us again take a logarithm on both sides
of the equations, and multiply by $i$:
\begin{eqnarray}\label{alllooplog}
2 L\,\arctan(2\,u_k)+i\,L\,\log
\left(\frac{1+g^2/2(x_k^-)^2}{1+g^2/2(x_k^+)^2}\right)
&=&
2\,\pi\,\tilde n_k-
2\,\sum_{\textstyle\atopfrac{j=-s/2}{j\neq 0}}^{s/2}
\arctan\left(u_k-u_j\right) \cr
&+ &2\,i\,\sum_{\textstyle\atopfrac{j=-s/2}{j\neq 0}}^{s/2} \log
\left(\frac{1-g^2/2x_k^+x_j^-}{1-g^2/2x_k^-x_j^+}\right)
\end{eqnarray}
Here we have also relabeled the $s$ roots $u_k$ such that the index $k$
runs over the set $k=\pm1,\pm2, \ldots,\pm \frac{s}{2}$.
We have furthermore chosen, for convenience, to employ a different
choice for the branches of the logarithms as compared to \eqref{1looplog}.
Whereas in \eqref{1looplog} the branchcuts run through $u_k=0$ and
$u_k=u_j$, in our
alternative choice in \eqref{alllooplog} the arctan functions are analytic
at $u_k=0$ and $u_k=u_j$.
This replaces the ``bosonic'' mode numbers $n_k$ of \eqref{1looplog} by
``fermionic'' mode numbers $\tilde n_k$. For the lowest state (the only
one for $L=2$) we have, for even $s$,
\begin{equation}\label{fermimodes}
\tilde n_k=k +\frac{L-3}{2}\,\epsilon(k)
\qquad {\rm for} \qquad
k=\pm1,\pm2, \ldots,\pm \frac{s}{2}\, .
\end{equation}
To avoid confusion: We are still focusing on the same states, and just
chose to change the description.

Let us now proceed in close similarity
to the computation of the thermodynamic antiferromagnetic
ground state of the
Heisenberg magnet (see e.g.~\cite{faddeev}). In order to have a
uniform spacing between the indices of all roots it is convenient
to define $k'=k-\epsilon(k)/2$ such that
\begin{equation}\label{fermimodes2}
\tilde n_k=k' +\frac{L-2}{2}\,\epsilon(k)
\qquad {\rm for} \qquad
k'=\pm\frac{1}{2},\pm\frac{3}{2}, \ldots,\pm \frac{s-1}{2}\, .
\end{equation}
As $s \rightarrow \infty$ we introduce a smooth
continuum variable $x=\frac{k'}{s}$. The
excitation density may now be defined as $\rho(u)=\frac{d x}{d u}$.
We divide \eqref{alllooplog} by $s$, use \eqref{fermimodes2},
replace the sums by integrals, and, finally, take a derivative w.r.t.~$u$.
Note that we do {\it not} rescale $u$ by $1/s$.
Then \eqref{alllooplog} becomes
\begin{eqnarray}\label{allloopintegral}
\frac{L}{s}\,\frac{1}{u^2+\frac{1}{4}}&+&\frac{i\,L}{s}\frac{d}{du}\log
\left(\frac{1+g^2/2(x^-(u))^2}{1+g^2/2(x^+(u))^2}\right)
= \cr
&= &2\,\pi\,\rho(u)+\frac{2\,\pi}{s}\,(L-2)\,\delta(u)-
2\int_{-b}^bdu'\frac{\rho(u')}{(u-u')^2+1} \cr
& &~~~~~+2\,i\,\int_{-b}^b du'\,\rho(u')\,\frac{d}{du}\,\log
\left(\frac{1-g^2/2x^+(u)x^-(u')}{1-g^2/2x^-(u)x^+(u')}\right)\, .
\end{eqnarray}
It is convenient to split the density $\rho(u)$ into a one-loop
piece $\rho_0(u)$ and a higher-loop piece $\tilde \sigma(u)$:
$\rho(u)=\rho_0(u)+g^2\,\tilde \sigma(u)$.
Let us, accordingly, also split off from
\eqref{allloopintegral} the one-loop contribution
\begin{equation}\label{1loopintegral}
\frac{L}{s}\,\frac{1}{u^2+\frac{1}{4}}
=2\,\pi\,\rho_0(u)+\frac{2\,\pi}{s}\,(L-2)\,\delta(u)-
2\int_{-\frac{s}{2}}^{\frac{s}{2}}du'\frac{\rho_0(u')}{(u-u')^2+1}\, ,
\end{equation}
while the higher (two and beyond) loop part of \eqref{allloopintegral}
becomes
\begin{eqnarray}\label{2loopintegral}
0&=&
2\,\pi\,\tilde\sigma(u) \cr
& &
-2\int_{-\infty}^\infty du'\frac{\tilde \sigma(u')}{(u-u')^2+1} \cr
& &+\frac{2\,i}{g^2}\,\int_{-\frac{s}{2}}^{\frac{s}{2}}
du'\,\rho_0(u')\,\frac{d}{du}\,\log
\left(\frac{1-g^2/2x^+(u)x^-(u')}{1-g^2/2x^-(u)x^+(u')}\right) \cr
& &+2\,i\,\int_{-\infty}^{\infty} du'\,
\tilde \sigma(u')\,\frac{d}{du}\,\log
\left(\frac{1-g^2/2x^+(u)x^-(u')}{1-g^2/2x^-(u)x^+(u')}\right)\, .
\end{eqnarray}
We have dropped the second term on the l.h.s.~of \eqref{allloopintegral},
as it is easily seen to be suppressed to leading order in the large
$s$ limit. This reflects the independence of the large
$s$ scaling behavior of the lowest state on the twist $L$
even beyond the one-loop approximation, as long as $L\ll s$.
We have also extended the range of integration of the second and fourth
integral in \eqref{2loopintegral} from $\pm s/2$ to $\pm \infty$, to be
justified below.

As a consistency check of our procedure let us rederive the
one-loop solution of the previous section from \eqref{1loopintegral}.
There we used rescaled variables $\bar u=\frac{u}{s}$, and a
rescaled density $\bar \rho_0(\bar u)=s\,\rho_0(u)$ such that
$d\bar u \,\rho_0(\bar u)=du\, \rho_0(u)$.
Using the large $s$ expansions
\begin{eqnarray}
\frac{1}{2\,s}\,\frac{1}{\bar u^2+\frac{1}{4\,s^2}}&=&
\pi\,\delta(\bar u)
+\cO(\frac{1}{s})\, , \\
\frac{1}{s}\,\frac{1}{(\bar u-\bar u')^2+\frac{1}{s^2}}&=&
\pi\,\delta(\bar u-\bar u')
+\frac{1}{s}\,\frac{{\cal P}}{(\bar u-\bar u')^2}+
\cO(\frac{1}{s^2})\, ,
\end{eqnarray}
where ${\cal P}$ indicates a principal part, we find from
\eqref{1loopintegral}
\begin{equation}\label{1loopsingular2}
0=4\,\pi\,
\delta(\bar u)+2\,\pint_{-\bar b}^{\bar b} d\bar u'\,
\frac{\bar \rho_0(\bar u')}{(\bar u-\bar u')^2}\, ,
\end{equation}
which is, since $\epsilon'(\bar u)=2\,\delta(\bar u)$, precisely the
derivative of the one-loop singular integral equation
\eqref{1loopsingular}. Note that the $L$ dependence has indeed again
dropped out. We therefore find the same one-loop result
as in the previous section. It should be stressed that, even though
the kernel in \eqref{1loopintegral} is of difference form, and
the interval boundary values tend to $\pm \infty$, it is
{\it incorrect} to solve this equation by naive Fourier techniques.

Luckily, however, applying a Fourier transform leads to progress with
the {\it higher-loop} equation \eqref{2loopintegral}. The reason is that
the higher loop density fluctuations $\tilde \sigma(u)$ are
concentrated in the vicinity of $u=0$, i.e.~$\tilde \sigma(u) \neq 0$
iff $|u| \ll s/2$. This may be verified for twist $L=2$
operators by using the exact one-loop solution of appendix
\ref{app:hahn}, and numerically solving the linear problem
of computing the higher-loop corrections to the roots of
the Hahn polynomials from the Bethe equations \eqref{allloopbethe}.
We were thus indeed entitled to replace the integral boundaries
$\pm s/2$ by $\pm \infty$ in the second and fourth term on the
r.h.s.~of \eqref{2loopintegral}. The ``scale'' of the fluctuations
$\tilde \sigma(u)$ is set by the third term on the
r.h.s.~of \eqref{2loopintegral}. Let us calculate it,
using $\rho_0(u)=\bar \rho_0(\bar u)/s$, with $\bar \rho_0(\bar u)$
given by \eqref{1loopdenssol}:
\begin{eqnarray}
&& \frac{2\,i}{g^2}\,\int_{-\frac{s}{2}}^{\frac{s}{2}}
du'\,\rho_0(u')\,\frac{\partial }{\partial u}\,\log
\left(\frac{1-g^2/2x^+(u)x^-(u')}{1-g^2/2x^-(u)x^+(u')}\right)= \cr
& = & - \frac{2\,i}{g^2}\,
\sum_{r=1}^\infty\frac{1}{r}\,\left(\frac{g^2}{2}\right)^r\,
\int_{-\frac{s}{2}}^{\frac{s}{2}}
du'\,\rho_0(u')\,\frac{\partial}{\partial u}\,
\left[\frac{1}{x^+(u)^r}\frac{1}{x^-(u')^r}-
\frac{1}{x^-(u)^r}\frac{1}{x^+(u')^r}\right]= \cr
&=& \frac{E_0}{s}\,\left(\frac{1}{2}\frac{d}{du}\right)\,\left[
\frac{1}{x^+(u)}+\frac{1}{x^-(u)}\right]
+\ldots\, ,
\end{eqnarray}
where we have only kept the leading contribution. Note that
only the first, $r=1$ term in the expansion of the logarithm
contributes to this result, and we have used, cf.~\eqref{resolvent},
\eqref{1loopeng}, the relation
\begin{equation}
\int_{-\frac{s}{2}}^{\frac{s}{2}}du'\,\rho_0(u')\,\frac{1}{x^\pm(u')}
= \int_{-\frac{s}{2}}^{\frac{s}{2}}du'\,
\rho_0(u')\,\frac{1}{u'\pm\frac{i}{2}} + \ldots =
\frac{1}{s}\,G\left(\frac{\mp \, i}{2\,s}\right) + \ldots =
\frac{\mp \, i}{2\,s}\, E_0 + \ldots \, ,
\end{equation}
which is valid to leading order at large $s$.
It is now clear from \eqref{1loopengscaling} that $E_0/s\simeq 4\,\log(s)/s$
sets the scale of the density fluctuation
$\tilde \sigma(u)$ in \eqref{2loopintegral}.
We therefore define $\tilde \sigma(u)= - (E_0/s) \, \sigma(u)$,
i.e.
\begin{equation}
\rho(u)=\rho_0(u)-g^2\,\frac{E_0}{s}\,\sigma(u)\, .
\end{equation}
To this leading order, the density fluctuation
does not change the density normalization $\int_{-b}^bdu\,\rho(u)=1$,
i.e.~$\int_{-b}^bdu\,\rho_0(u)= 1 + \ldots$ since
$\underset{s \rightarrow \infty}{\lim}\,E_0/s=0$, see \eqref{1loopengscaling}.
Then \eqref{2loopintegral} becomes
\begin{eqnarray}\label{2loopintegral2}
0&=&
2\,\pi\,\sigma(u) \cr
& &-2\int_{-\infty}^\infty du'\frac{\sigma(u')}{(u-u')^2+1} \cr
& &-\left(\frac{1}{2}\frac{d}{du}\right)\,\left[
\frac{1}{x^+(u)}+\frac{1}{x^-(u)}\right] \cr
& &+2\,i\,\int_{-\infty}^{\infty} du'\,
\sigma(u')\,\frac{\partial}{\partial u}\,\log
\left(\frac{1-g^2/2x^+(u)x^-(u')}{1-g^2/2x^-(u)x^+(u')}\right)\, .
\end{eqnarray}
We now introduce the Fourier transform $\hat \sigma(t)$ of the
fluctuation density $\sigma(u)$
\begin{equation}\label{fourier}
\hat \sigma(t)=e^{-\frac{t}{2}}\,
\int_{-\infty}^\infty du\,e^{-i t u}\,\sigma(u)\, ,
\end{equation}
where we have also included a factor $e^{-\frac{t}{2}}$ for
notational convenience. Fourier transforming
$e^{-\frac{t}{2}}\,\int_{-\infty}^\infty du\,e^{i t u}\,\times$
equation \eqref{2loopintegral2} we find, after some calculation
(see appendix \ref{app:fourier}),
\begin{equation}\label{prefredholm}
0=2 \pi\,\hat \sigma(t)-2 \pi\,e^{-t}\,\hat \sigma(t) -2
\pi\,e^{-t}\,\frac{J_1(\sqrt{2}\,g\,t)}{\sqrt{2}\,g} +4
\pi\,g^2\,t\,e^{-t} \int_0^{\infty}dt'\, \hat
K(\sqrt{2}\,g\,t,\sqrt{2}\,g\,t')\,\hat \sigma(t')\, ,
\end{equation}
where the four terms in \eqref{prefredholm} correspond, respectively,
to the four terms in \eqref{2loopintegral2}, and the kernel $\hat K$
is given in terms of Bessel functions by
\begin{equation}\label{kernel2}
\hat K(\sqrt{2}\,g\,t,\sqrt{2}\,g\,t') \, = \frac{1}{\sqrt{2}\,g}\,
\, \frac{J_1(\sqrt{2}\,g\,t) \, J_0(\sqrt{2}\,g\,t') \, -
\, J_0(\sqrt{2}\,g\,t) \, J_1(\sqrt{2}\,g\,t')}
{t \, - \, t'}\, .
\end{equation}
Note that the Fourier transform only diagonalizes the ``main''
scattering term in \eqref{2loopintegral2}, i.e.~the kernel
$1/((u-u')^2+1)$. So we are still left with an integral equation.
However, the higher-loop equation \eqref{prefredholm} is,
in view of \eqref{kernel2}, and in contradistinction to
the one-loop equation \eqref{1loopsingular}, {\it non-singular}.
It may be rewritten in the form \eqref{fredholm} stated in
the introduction. Finally, the all-loop energy is found from
\eqref{allloopmomeng} to be
\begin{eqnarray}
E(g) & = & s\,\int_{-\frac{s}{2}}^{\frac{s}{2}}\,du\,\rho(u)\,
\left(\frac{i}{x^+(u)}-\frac{i}{x^-(u)}\right) + \ldots
\label{eneru} \\ & = &
E_0-g^2\,E_0\,\int_{-\infty}^{\infty}\,du\,\sigma(u)\,
\left(\frac{i}{x^+(u)}-\frac{i}{x^-(u)}\right) + \ldots \nonumber
\end{eqnarray}
to leading order in $s$.
In terms of the Fourier transformed density $\hat \sigma(t)$,
cf.~\eqref{fourier}, this becomes (see again appendix \ref{app:fourier})
\begin{equation}\label{allloopengscaling}
E(g) = E_0\left(1 - 4 \, g^2 \, \int_0^\infty \,
dt\, \hat \sigma(t)\,
\frac{J_1(\sqrt{2}\, g\, t)}{\sqrt{2}\, g \, t}\right) + \ldots
\end{equation}
with $E_0 = 4 \,\log(s) + \ldots \, $.
Notice that, in line with general expectations, we have
just shown that the Bethe ansatz of \cite{BS2}
indeed leads to the logarithmic scaling behavior \eqref{log}
{\it to all orders in perturbation theory},
in agreement with general expectations, see e.g.~the discussions
in \cite{KLOV},\cite{korchemsky},\cite{korchemsky2}.
In view of \eqref{log},\eqref{Delta},\eqref{1loopengscaling}
this indeed yields our proposed conjecture for the all-loop
scaling function $f(g)$ announced in \eqref{scalingfunction}.
The proposed scaling function as found from the Bethe ansatz
possesses further remarkable properties, to which we will
now turn our attention.

\subsection{Weak-Coupling Expansion and Transcendentality}
\label{weakcoupling}

The Fredholm form of the higher-loop integral equation \eqref{fredholm}
or \eqref{prefredholm} is ideally suited for the explicit perturbative
expansion of the scaling function $f(g)$ of \eqref{log} to high orders.
Both the inhomogeneous, first term as well as the kernel of \eqref{fredholm}
have a regular expansion in even powers of $g$ around $g=0$.
We may therefore also expand the transformed density $\hat \sigma(t)$
in even powers of $g$ and solve \eqref{fredholm} iteratively
\begin{equation}
\hat \sigma(t)= \frac{1}{2} \, \frac{t}{e^t \, - \, 1}
-g^2\,\left(\frac{1}{8} \, \frac{t^3}{e^t \, - \, 1} \, + \,
\frac{1}{2} \, \zeta(2) \, \frac{t}{e^t \, - \, 1}\right)+\ldots\,
, \label{pertsigmat}
\end{equation}
where we have used the following representation of the
Riemann zeta function:
\begin{equation}\label{zeta}
\zeta(n+1) =
\frac{1}{n!}\,\int_0^\infty \frac{dt \; t^n}{e^t \, - \, 1}\, .
\end{equation}
Furthermore, the expression for the
scaling function \eqref{scalingfunction} may also be expanded
in a Taylor series in $g^2$:
\begin{equation}
f(g) = 4\,g^2  - 4\,g^4\, \int_0^\infty
\frac{dt \; t}{e^t \, - \, 1} \, + g^6\, \Bigl(2\, \int_0^
\infty \frac{dt \; t^3}{e^t \, - \, 1} \, + 4\, \zeta(2) \,
\int_0^\infty \frac{dt \; t}{e^t \, - \, 1} \, \Bigr) \, + \, \ldots \, .
\end{equation}
We again use \eqref{zeta} and we find to e.g.~six-loop order
\begin{eqnarray}\label{longprediction1}
f(g) & = & 4 \, g^2 \, - 4 \, \zeta(2) \, g^4 \, +  \,
\Bigl(4 \, \zeta(2)^2 \, + \, 12 \, \zeta(4)\Bigr) \, g^6 \\ && - \, \Bigl(4
\, \zeta(2)^3+ 24 \, \zeta(2) \zeta (4)-4 \, \zeta (3)^2+50 \, \zeta (6)
\Bigr) \, g^8 \nonumber \\ && + \, \Bigl(4 \, \zeta (2)^4+36 \,
    \zeta (2)^2 \zeta(4)-8 \, \zeta (2) \zeta (3)^2+100 \, \zeta (2) \zeta (6)
\nonumber \\ && \phantom{+ \, \Bigl(} -40 \, \zeta (3) \zeta (5)+39 \,
    \zeta (4)^2 + 245 \, \zeta (8) \Bigr) \, g^{10}
\nonumber \\ && - \,
    \Bigl(4 \, \zeta (2)^5+48 \, \zeta (2)^3 \zeta
    (4)-12 \, \zeta (2)^2 \zeta (3)^2+150 \, \zeta (2)^2 \zeta (6)
\nonumber \\ && \phantom{+ \Bigl(} -80 \, \zeta (2) \zeta (3) \zeta (5) +
114 \, \zeta (2) \zeta (4)^2 + 490 \, \zeta (2) \zeta (8)-18 \,
\zeta (3)^2 \zeta (4)  \nonumber \\ && \phantom{+ \Bigl(} -210 \, \zeta (3)
\zeta (7) +345 \, \zeta (4) \zeta (6)- 102 \, \zeta (5)^2 +
1323 \, \zeta (10) \Bigr) \, g^{12} \, + \, \ldots \nonumber
\end{eqnarray}
It is easy to go to much higher orders if desired (we have expanded to
20-loop order $g^{40}$). It is seen that the $\ell$-loop
$\cO(g^{2 \ell})$ contribution to the anomalous dimension is a
sum of products of zeta functions.
What is more, the arguments of the zeta functions of each product
always add up to the number $2\,\ell-2$. This is a test of the
{\it ``transcendentality principle''} of
Kotikov, Lipatov, Onishchenko and Velizhanin as spelled out in
\cite{KL,KLV,KLOV}, and we see that our Bethe ansatz is {\it consistent}
with this principle\footnote{
To be more precise, here we have tested a weaker form of the
transcendentality principle of \cite{KLOV}. The stronger form
applies to the finite $s$ case, and states that the indices of
certain harmonic sums add up to $2\,\ell-1$. We suspect that
our all-loop Bethe ansatz is also consistent with the stronger
version, see also \cite{S}. 
Our finding certainly supports
this, as the weaker principle is a consequence of the
stronger one. It would be exciting to fully prove the latter
from the $L=2$ finite $s$ Bethe equations
\eqref{allloopbethe},\eqref{allloopmomeng}.
}.
Finally it is also seen that the numerical
coefficients in front of each zeta function product are
integers\footnote{
Actually, with our convention \eqref{convention},
higher terms beyond the order we have printed in \eqref{longprediction1}
develop powers of 2 in the denominator.
We however checked up to order $g^{40}$ that our scheme
yields indeed integer numbers in front of the zeta-functions if
$g$ is rescaled as $g \, \rightarrow \, \sqrt{2} \, g$, 
which is Lipatov's et.al.~convention. 
}.

Note that the expansion \eqref{longprediction1} may be written more
compactly when expressing the zeta functions of even arguments
through powers of $\pi$ times rational numbers:
\begin{eqnarray}\label{longprediction2}
   f(g) & = & 4 \, g^2  \\
   && -\frac{2}{3} \, \pi ^2 \, g^4 \nonumber \\
   && + \frac{11}{45} \, \pi ^4 \, g^6 \nonumber \\
   && - \Bigl( \frac{73}{630} \, \pi ^6 - 4 \, \zeta (3)^2 \Bigr) \,
g^8 \nonumber \\
   && + \Bigl(\frac{887}{14175} \, \pi ^8
-\frac{4}{3} \, \pi ^2 \, \zeta (3)^2
-40 \, \zeta (3) \zeta (5) \Bigr)  \, g^{10} \nonumber \\
   && - \Bigl(\frac{136883}{3742200} \, \pi ^{10}-\frac{8}{15} \, \pi ^4
    \, \zeta (3)^2-\frac{40}{3} \, \pi ^2 \, \zeta (3) \zeta (5) \nonumber \\
   && \phantom{-\Bigl(} \; -210 \, \zeta (3) \zeta (7) -
   102 \, \zeta (5)^2 \Bigr)\, g^{12} \nonumber \\
&& + \, \ldots .\nonumber
\end{eqnarray}
Clearly each $\pi$ contributes one ``unit'' of transcendentality.
This however obscures the integer nature of the numerical coefficients
(c.f.~footnote 8).

It is instructive to investigate whether the (BMN scaling-preserving) ``AFS'' dressing factor \cite{AFS,S,BS2} for the 
(approximate, see \cite{SZZ,BT,SZ})
{\it string}\footnote{
Our motivation here is not
so much string theory as such (in particular we investigate the
dressing factor at weak coupling, while its original design
demands strong coupling) but rather the fact that this type
of dressing factors are known to naturally appear in certain variant,
asymptotically integrable spin chains \cite{BK}. While these
studies were done for compact magnets, it is likely that they
may be generalized to the non-compact case of interest in this paper.
The variant models tend to violate the Feynman rules of the gauge
field theory, which is our main motivation for investigating whether they
preserve the transcendentality principle.
}
Bethe ansatz 
\eqref{allloopbethe},\eqref{allloopmomeng} is compatible with
the transcendentality principle. 
Possible (BMN scaling-violating)
gauge dressing factors are briefly treated in the next section \ref{breakdown}.

The AFS ansatz leads to a modification, at three
loops and beyond, of the integral equation \eqref{fredholm}
\begin{equation}\label{fredholm2}
\hat \sigma(t) \,
= \, \frac{t}{e^t - 1} \, \Bigl[ K'(\sqrt{2} g \, t, \, 0) \, -
\, 2 \, g^2 \int_0^\infty dt' \,
K'(\sqrt{2} g \, t, \, \sqrt{2} g \, t') \, ,
\tilde \sigma(t') \,
\Bigr]\, ,
\end{equation}
where the modified kernel $K'$, see appendix \ref{app:fourier}, reads
\begin{equation}\label{extra}
K'(\sqrt{2} g \, t, \, \sqrt{2} g \, t') \, = \, \hat K(\sqrt{2} g \, t,
\, \sqrt{2} g \, t') \, + \, \sqrt{2} g \, \tilde K(
\sqrt{2} g \, t, \, \sqrt{2} g \, t')\, ,
\end{equation}
with
\begin{equation}\label{ksigdef2}
\tilde K(t,t') \, = \, \frac{t (J_2(t) \, J_0(t') - J_0(t) \,
J_2(t'))} {t^2 - {t'}^2} \, , \qquad \tilde K(t,0) \, = \,
\frac{J_2(t)}{t} \, .
\end{equation}
The dressing factor then modifies the scaling function
$f(g) \rightarrow f(g)+\delta f(g)$ in the following fashion:
\begin{eqnarray}
\delta f(g) & = & 0 \times g^2 \\ &&
0 \times g^4 \nonumber \\ && -4 \, \zeta (3)  \,
g^6 \nonumber \\ && + \Bigl(\frac{4}{3} \, \pi ^2 \, \zeta (3)+
20 \, \zeta(5) \Bigr)  \, g^8 \nonumber \\
   && - \Bigl(\frac{23}{45} \, \pi ^4 \, \zeta (3)+ \frac{20}{3} \, \pi ^2 \,
    \zeta (5)+105 \, \zeta (7)-4 \, \zeta (3)^2  \Bigr)  \, g^{10}
\nonumber \\
   && + \Bigl(\frac{71}{315} \, \pi ^6 \, \zeta (3)+\frac{79}{30} \,
\pi ^4 \,
\zeta (5)+35 \pi ^2 \, \zeta (7)-8 \, \zeta (3)^3 + 588 \, \zeta (9)
\nonumber
\\  && \phantom{+\Bigl(} \; -2 \, \pi ^2 \, \zeta
    (3)^2-36 \, \zeta (3) \zeta
    (5) \Bigr) \, g^{12} \nonumber \\
&& + \, \ldots \nonumber
\end{eqnarray}

We see that the integrable modification of the long-range Bethe
ansatz of \cite{BS2} by the
``stringy'' AFS \cite{AFS} 
dressing factor {\it violates} the
transcendentality principle\footnote{For the gauge theory ansatz
the transcendentality principle is a consequence of scaling: the
arguments of potential and kernel in \eqref{prefredholm} are
$\sqrt{2} g \, t, \, \sqrt{2} g \, t'$ so that the order in $g$ is
linked to the total power of $t$ and $t'$ which defines the level
of transcendentality. The string theory ansatz \eqref{fredholm2}
breaks the pattern only because of the presence of the extra
$\sqrt{2} g$ in front of $K'$ in equation \eqref{extra}. Initially
this introduces a mismatch by one unit; by iteration the effect
fans out higher up in the perturbative expansion.}, as now the
arguments of the Riemann zeta functions no longer add up to
$2\,\ell-2$.


\subsection{Breakdown of BMN scaling and the Scaling Function}
\label{breakdown}

Here we will demonstrate interesting
connections between {\it BMN scaling} \cite{bmn} on the one hand and
our Bethe ansatz method for the scaling function on the other. 
It is by now rather firmly
established that BMN scaling in perturbative gauge theory
can only break down, at four loops or beyond, through a dressing factor 
of the general type just discussed, see in particular 
\cite{BK},\cite{Sproofs}.
This happens in e.g.~the plane-wave matrix model, see \cite{FKP}.

Let us sketch the quantitative derivation of this effect, restricting
ourselves for simplicity to four loops, where its detection might
still be within reasonable reach of sophisticated field theory methods, 
maybe along the lines of \cite{BerDixSmi}.

The first modification of the asymptotic Bethe equations of \cite{BS2}
which is still consistent with current knowledge on the integrable
structure of $\cN=4$ gauge theory would lead to the
following correction of the higher loop Bethe equations 
\eqref{allloopbethe}
\begin{equation}\label{allloopbethemod}
\left(\frac{x^+_k}{x^-_k}\right)^L=
\prod_{\textstyle\atopfrac{j=1}{j\neq k}}^s
\frac{x_k^--x_j^+}{x_k^+-x_j^-}\,
\frac{1-g^2/2x_k^+x_j^-}{1-g^2/2x_k^-x_j^+}\,
\sigma^2(u_k,u_j)
\end{equation}
with
\begin{equation}
\sigma^2(u_k,u_j)=
e^{i\,\beta\,g^6\,\left(q_2(u_k)\, q_3(u_j)-q_3(u_k)\, q_2(u_j) \right)
+ \ldots}\, ,
\end{equation}
see \cite{AFS,BS2,BK,BT,Sproofs} for details\footnote{
The detailed argumentation which allows to draw this conclusion is 
actually rather subtle and requires putting together various results.
The main steps are: (1) The three-loop Bethe ansatz is solidly known.
(2) The structure of the four-loop Bethe ansatz is also known, up to the 
term involving $\beta$ in \eqref{allloopbethemod}, in the $\alg{su}(2)$
sector \cite{BK}. 
(3) The multiplicative modification affecting the $\alg{su}(2)$ 
sector as in \eqref{allloopbethemod} must also multiplicatively affect
in the same fashion the $\alg{sl}(2)$ sector, as first conjectured
in \cite{S} and later proved in \cite{Sproofs}.},
and the definition of the charges $q_r(u)$.
The dots indicate further terms which might affect five loops and higher.

A non-zero value for $\beta$ leads to ``soft-breaking'' of BMN scaling: The
two-excitation problem can be solved exactly \cite{MZ}, because the momentum
constraint implies $u_2 \, = \, - u_1$. For spin chain length $L=J$,
where $J$ is the BMN R-charge, the
different states are distinguished by the Bethe roots $u_{1,n} \, =
\, \frac{1}{2} \cot\left(\frac{\pi n}{J+1}\right)$. The higher order
corrections similarly come out in terms of trigonometric functions.
The string spectrum $\Delta-J$ is reproduced by Taylor expanding in 
$1/J$ when $n$ is small:
\begin{equation}
\Delta-J = 2+8 \, g^2 \left(\frac{n \pi}{J}\right)^2 - 16 \, g^4
\left(\frac{n \pi}{J}\right)^4 + 64 \, g^6 \left(\frac{n \pi}{J}\right)^6
 - 320 \, g^8 \left(\frac{n \pi}{J}\right)^8 - 512 \, g^8 \frac{\beta}{J}
\left(\frac{n \pi}{J}\right)^6 + \ldots
\end{equation}
The dots stand for higher orders in $g^2$ and, at any given order,
terms subleading in $1/J$.
We see the emergence of the effective coupling constant $g^2/J^2$ in the
first four terms of the last formula, while the last term has
$g^8/J^7$, so that it diverges in the BMN limit
$g,J \, \rightarrow \, \infty$ with $g/J$ fixed.

The modified Bethe ansatz \eqref{allloopbethemod} requires
replacing the kernel in \eqref{kernel2} by
\begin{equation}
\hat K(\sqrt{2} g \, |t|, \, \sqrt{2} g \, |t'|) 
\rightarrow
\hat K(\sqrt{2} g \, |t|, \, \sqrt{2} g \, |t'|) \,+
2 \, \beta \, (\sqrt{2} g) \,
\frac{J_2(\sqrt{2}\,g\,|t|)\,J_1(\sqrt{2}\,g\,|t'|)}{|t\,t'|}\,
+\ldots\, \, .
\end{equation}
The third term on the r.h.s.~of \eqref{2loopintegral2} becomes
\begin{equation}
-\left(\frac{1}{2}\frac{d}{du}\right)\,\left[
\frac{1}{x^+(u)}+\frac{1}{x^-(u)}\right] \, - \, \beta \, g^4 \, 
\frac{d}{du} \, q_3(u)\, +
\ldots\, ,
\end{equation}
or, after Fourier transforming (c.f.~third term in \eqref{prefredholm})
\begin{equation}
 -2 \pi\,e^{-t}\,\left( \frac{J_1(\sqrt{2}\,g\,t)}{\sqrt{2}\,g} \, + \, 
2 \, \beta \, g^2 \, J_2(\sqrt{2}\,g\,t) \, \right) \, +
\ldots\, .
\end{equation}
This modifies the four-loop $\cO(g^8)$ term of the 
scaling function \eqref{longprediction2} to
\begin{equation}
- \left( \frac{73}{630}\,\pi^6-4\,\zeta(3)^2 + 8 \, \beta \, \zeta(3)
\right) \, .
\end{equation}
Note that transcendentality is violated unless $\beta$ is a 
rational number times $\zeta(3)$ (or $\pi^3$). 
A particularly curious case would
be $\beta=\frac{1}{2}\zeta(3)$, which would lead to the much
simpler four-loop answer $-\frac{73}{630}\,\pi^6$.

Note that such a modified Bethe ansatz would also change the anomalous
dimensions of all operators in other sectors. E.g.~in the $\alg{su}(2)$
sector we would find for the length $L=5$ 
operator $\Tr X^2\,Z^3 + \ldots$ (this case is actually equivalent
to the $\alg{sl}(2)$ twist three operator $\Tr D^2\,Z^3 + \ldots$)
to four loops
\begin{equation}
E(g)=4-6\,g^2+17\,g^4-(\frac{115}{2}+8\,\beta)\,g^6\, + \ldots \, .
\end{equation}
It would be very interesting if the modification were
non-rational\footnote{Clearly the so far proposed Bethe ans\"atze \cite{BS2}
also lead to a transcendentality principle at weak coupling:
If we assign, in accordance with the meaning of the word,
transcendentality degree zero to rational or algebraic numbers,
then weak coupling dimensions of operators carrying finite
charges (i.e.~without taking limits of large R-charges or large
spin quantum numbers) are always of zero degree in the 
currently proposed ans\"atze.
On the other hand, zeta functions do appear naturally in individual
higher loop Feynman diagrams, and, from this point of view,
might well appear in high order contributions to
anomalous dimensions. }.
Incidentally, we see that $\beta \neq 0$ would also rule out
the Hubbard Hamiltonian as a candidate for the $\alg{su}(2)$
dilatation operator beyond three-loop order, c.f.~eq.(68) in \cite{RSS}.


\subsection{Strong-Coupling Expansion and String Theory}
\label{strongsec}

The Fourier-transformed integral equation \eqref{prefredholm} does
not lend itself to strong coupling analysis due to the oscillatory
nature of the kernel \eqref{kernel2}. We rather return to the
configuration space integral equation \eqref{2loopintegral2}.

The two diagrams in Figure 1 give a series of plots of the root
density for progressively higher values of the coupling constant.
\begin{figure}[t]
\centerline{\includegraphics[width=6cm]{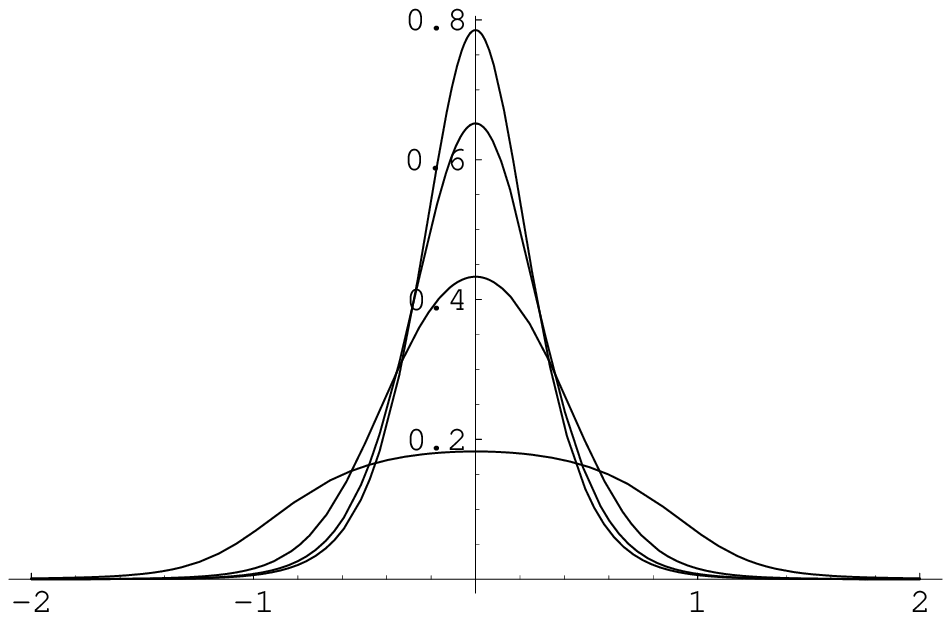}\qquad
\includegraphics[width=6cm]{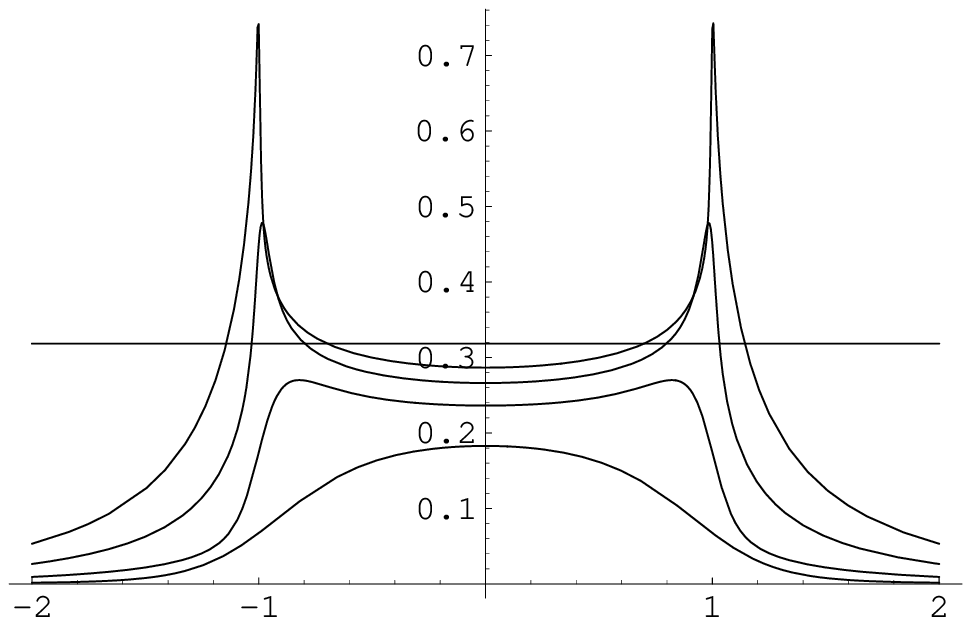}}
\caption{\label{fig1}\small The density of Bethe roots at weak
(left) and strong coupling (right).}
\end{figure}
The left picture shows the weak coupling regime; the graphs depict
the root density at $\sqrt{2} g \, = 0, \, 1/4, \, 1/2, \, 1$,
respectively. The $\sqrt{2} g \, = \, 0$ distribution is the
tallest peak. It is given by the Fourier back-transform of the
first term in \eqref{pertsigmat}:
\begin{equation}
\sigma_0(u) \, = \, \frac{\pi}{4} \, \frac{1}{\cosh^2(\pi \, u)}
\end{equation}
All other curves are numerical solutions of
\eqref{2loopintegral2}. Augmenting the coupling constant makes the
peak around $u \, = \, 0$ become wider and flatter.

In the second diagram we plotted $2 g^2 \, \sigma(u / (\sqrt{2}
g))$ for $\sqrt{2} g \, = \, 1, \, 4, \, 16, \, 64$. With
increasing coupling the graphs rise; they develop peaks at $\pm 1$
while the middle parts tend to $1/\pi$. On undoing the scaling we
would nevertheless recover the tendency seen at weak coupling,
i.e. the support of the root density roughly stretches to the
interval $[-\sqrt{2} g, \, \sqrt{2} g]$ within which the density
tends to
\begin{equation}
\sigma_{\infty}(u) \, = \, \frac{1}{2 \pi g^2} \, .
\end{equation}
Note that the constant function $\sigma(u) \, = \, 1/(2 \pi g^2)$
is an exact solution of \eqref{2loopintegral2} if the support is
extended to the entire real axis (likewise $\sigma(t) =
\delta(t)/g^2$ is a solution of \eqref{prefredholm}). Furthermore,
$\sigma_{\infty}(u)$ would exactly cancel the leading $O(g^2)$
contribution to the scaling function \eqref{allloopengscaling},
thus yielding the $O(g)$ asymptotics expected from string theory.

Numerically, we could confirm the cancellation of the $O(g^2)$
part of the scaling function up to an error of a few per cent, but
reliable predictions for subleading terms remained out of reach.
It is indispensable to understand the strong coupling regime by
analytic means. We hope to clarify the issue in future work.


\section*{Acknowledgments}

We would like to thank Niklas Beisert, Zvi Bern, Lance Dixon,
Sergey Frolov, David Gross, Volodya Kazakov, Valery Khoze,
David Kosower, Anatoli Kotikov, Stefano Kovacs, Lev Lipatov,
Juan Maldacena, Andrei Onishchenko, Yaron Oz, Jan Plefka,
Joe Polchinski, Adam Rej, Sakura Sch\"afer-Nameki, Didina Serban,
Emery Sokatchev, Marija Zamaklar and Kostya Zarembo for their interest and for
helpful discussions.
The work of Burkhard Eden was supported by the
{\it Deutsche Forschungsgemeinschaft}, Schwerpunktprogramm (1096)
{\it ``Stringtheorie im Kontext von Teilchenphysik,
Quantenfeldtheorie, Quantengravitation, Kosmologie und Mathematik}.


\appendix


\section{Two-point Functions in the $sl(2)$ Sector}
\label{app:twopoint}

\subsection{Perturbative CFT in the Dimensional Reduction Scheme}

We shall restrict our attention to leading $N$ (planar) two-point functions
of single trace operators in the $\alg{sl}(2)$ sector. For any given spin chain
with length=twist $L$ there are many distinct operators differing in the total
number of derivatives and their positioning on the sites of the chain.

Renormalization must be done in such a way as to un-mix these states and to
make their correlators finite. The theory is then seen to be conformally
invariant; for example the two-point function of a renormalized primary
operator of spin $s$ has the form \cite{pisa}
\begin{equation}
\langle P^s(1) \bar P^s(2) \rangle \, = \, \frac{c(g^2) J_{\mu_1 \nu_1}(x_{12})
\ldots J_{\mu_s \nu_s}(x_{12})}{(x_{12}^2)^{\Delta(g^2)}} \label{conf} \, ,
\end{equation}
where
\begin{equation}
J_{\mu \nu}(x) \, = \, \eta_{\mu \nu} - 2 \frac{x_\mu x_\nu}{x^2}
\end{equation}
is the inversion tensor, and the $\mu$ and $\nu$ indices are separately made
traceless and symmetric. Knowledge of any one term in the product
of inversion tensors is sufficient to reconstruct the full correlator.
In \cite{mejan} we considered the term with no $\eta$ symbol, because we were
interested in a minimal set of graphs (trace terms are potentially more
divergent and there are also a few Feynman diagrams which always carry at
least one power of $\eta$ on dimensional grounds). In the present work we wish
to construct the asymptotic two-loop dilatation operator in the $sl(2)$ sector.
The task is greatly simplified by focusing on the pure trace terms, because
these obviously cannot exist between operators of different spin. This property
becomes important when subtracting out disconnected parts.

As before, we use ${\cal N}=2$ superfields and regularize by SSDR
(supersymmetric dimensional reduction) \cite{DimRed} in $x$-space.
This amounts to doing the superalgebra as in four dimensions, while the
underlying scalar propagator is modified as in standard dimensional
regularization:
\begin{equation}
\langle Z(1) \bar Z(2) \rangle \, = \, \frac{c_0}{x_{12}^2} (\mu x_{12}^2)
^\epsilon \, , \qquad c_0 \, = \, - \frac{1}{4 \pi^2} \, , \qquad
\square_1 \langle Z(1) \bar Z(2) \rangle \, = \, \delta(x_{12}) \, ,
\label{ssdr}
\end{equation}
although we suppress the mass scale $\mu$ throughout the article.\footnote{
Our discussion of the renormalization of conformal correlators in $x$-space
using the SSDR scheme is built upon the works \cite{EJS,EJSS,mejan}.}
The tree-level correlators of operators of length $L$ and spin $s$ thus
contain the $x$-space structure
\begin{equation}
 X(L,s) \, = \, \frac{N^L \eta_{z \bar z}^s}
{(- 4 \pi^2)^L (x_{12}^2)^{L (1-\epsilon) + s}} \label{etachoice}
\end{equation}
and a whole series of terms with $x_{12}$ with open indices, which may be
recovered by appealing to conformal invariance.

In order to extract the one- and two-loop anomalous dimensions we must keep
track of the leading and sub-leading order in the $\epsilon$ expansion of the
bare correlators:
\begin{eqnarray}
\langle \, {\cal O}_i \, \bar {\cal O}_j \, \rangle & = &
X(L,s) \, \Bigl[ \, \bigl( T_{0 \, ij} \, + \, \epsilon \, T_{1 \, ij} \bigr)
\, + \, g^2 \, \bigl( A_{11 \, ij} \, \frac{1}{\epsilon} \, + \, A_{10 \, ij}
\bigr) \, (x_{12}^2)^\epsilon \nonumber \\ && \phantom{X(L,s) \, \Bigl[}
+ \, g^4 \, \bigl(A_{22 \, ij} \, \frac{1}{\epsilon^2} \, + \,
A_{21 \, ij} \, \frac{1}{\epsilon} \, + \, A_{20 \, ij} \bigr) \,
(x_{12}^2)^{2 \epsilon} \, + \, \ldots \, \Bigr]
\end{eqnarray}
where the Yang-Mills coupling constant is dressed by\footnote{We deviate
from the convention in \cite{mejan} by a coupling constant rescaling so as to
be more in line with the literature.}
\begin{equation}
g^2 \, = \, \frac{g^2_{YM} N}{8 \pi^2}
\end{equation}
and the fractional powers of $x_{12}^2$ arise from the integration measure
in the Feynman graphs defining the one- and two-loop contributions.

Consistency of ${\cal N}=4$ as a conformal field theory grants that
$T_0, \, A_{11}, \, A_{22}$ are simultaneously diagonalizable. In a
diagonal basis $\{O_i\}$ they obey
\begin{equation}
A_{11} \, = \, \Gamma_1 \, T_0 \, , \qquad A_{22} \, = \, \frac{1}{2} \,
\Gamma_1^2 \, T_0 \, .
\end{equation}
Here $\Gamma_1$ is also diagonal and contains the one-loop anomalous
dimensions $\gamma_{1 \, i}$.

The divergences are removed by introducing $Z$ matrices of the
form
\begin{equation}
Z \, = \, R \, + \, g^2 \, B \, + \, g^4 \, \bigl(C_1 \, \frac{1}
{\epsilon} \, + \, C_0 \bigr) \, + \ldots
\end{equation}
where $R$ is diagonal and has as its entries the $Z$-factors for the
individual operators
\begin{equation}
{\cal Z}_i \, = \, 1 \, + \, g^2 \, \frac{z_{11 \, i}}{2 \epsilon} \, +
\, g^4 \, \bigl( \frac{z_{22 \, i}}{4 \epsilon^2} \, + \,
\frac{z_{21 \, i}}{4 \epsilon} \bigr) \, + \, \ldots
\end{equation}
while $B,\, C$ have zero on the diagonal. The $Z$ factors and the anomalous
dimensions are determined from the bare two-point functions by imposing
\begin{equation}
F \, = \, Z \, \langle {\cal O} \, \bar {\cal O} \rangle \,
Z^\dagger \label{starthere}
\end{equation}
where $F$ is again diagonal and is defined by the renormalized two-point
functions
\begin{equation}
f_i \, = \, X(L,s)|_{\epsilon=0} \, (a_{0 \, i} \, + \, g^2 \,
a_{1 \, i} \, + \, g^4 \, a_{2 \, i}) \, (x_{12}^2)^{ - g^2
\, \gamma_{1 \, i} \, - \, g^4 \, \gamma_{2 \, i}} \, + \,
\ldots
\end{equation}
To be more precise, we demand that both sides be equal at each order in $g^2$
up to positive powers of $\epsilon$. The resulting system of equations does
not completely fix $C_1, \, C_0$, so that we limit our scope to the
determination of $R,B,a_{0i},a_{1i},\gamma_{1i},\gamma_{2i}$. We may thus
drop the constant part $A_{20}$ of the $g^4$ two-point functions from our
analysis.

\subsection{Graphs}

We exploit the ${\cal N}=2$ superfield formalism in order to minimize the
number of Feynman diagrams. For a quick review of the essentials of the
formalism and expressions for the graphs we would like to refer the reader to
\cite{mejan}, where two-loop two-point functions of operators of length three
are discussed. Our notations and conventions are in fact borrowed from that
work; in particular, the article contains a list of graphs upon which we draw
here. However, in \cite{mejan} the $(x_z x_{\bar z})^s$ term of the
two point functions was used, so that some graphs could be omitted because
they always come with $\eta_{z \bar z}$.

At order $g^2$, we additionally have to take into account a graph
$F$ (see Figure 3 below) in which a free vector line goes from the
connection in $D_z$ on the left end of the two-point function to
that in $D_{\bar z}$ on the right (the Feynman gauge vector
propagator is proportional to $\eta$). Correspondingly, there is
an $O(g^4)$ graph consisting of the same free line paired with the
divergent one-loop graph $G_0$. On the other hand, we do not need
to consider the combination of the free vector line with the
``BPS-like'' $O(g^2)$ integral $B_0$ since this configuration
stays finite. Next, in \cite{mejan} we could drop the product $G_0
\,* \,  B_0$ as $G_0$ only has a simple pole (in $x$-space) while
the part of $B_0$ without $\eta$ is a contact term also when there
are partial derivatives on the outer legs, i.e.~it is always
$O(\epsilon)$. Terms in $B_0$ which involve $\eta_{z \bar z}$ are
finite, i.e. O(1), so that in the present context the product $G_0
\, * \, B_0$ becomes relevant.

With respect to the genuine two-loop integrals there are not many changes: the
finiteness of some terms which we dropped from graph $G_3$ remains guaranteed
and hence we may take over the simplified sum $G_3 \, + \, G_4$ given in
formula (61) in \cite{mejan}. The ``BPS-like'' graphs behave in the
same manner as $B_0$: the part without $\eta$ is a contact term and the other
parts are finite. They can still safely be omitted.

A first difference is that graphs $G_{10}$ and $G_{11}$ start to contribute:
before, the poles from these graphs cancelled in the sum over all diagrams
within each class; this is not the case in the new situation.\footnote{We
point out an error in formula (55) in the original version of \cite{mejan}:
two parts of the integral were added with a wrong relative sign. The
cancellation of the associated poles in the calculation of \cite{mejan} can be
verified for both ``halves'' on their own so that the mistake did not show.
The correct expression for $G_{10}$ is:
\begin{equation}
G_{10} \, = \, (12) \bigl[ \,
\partial_{\nu 14} \partial_{\mu 23} + \eta_{\mu \nu} \square_{34} / 4 \, -
(1 - (12)(1^-2^-)) \partial_{\mu 13} \partial_{\nu 24} \, \bigr]
\end{equation}
}

But there are also three genuinely new graphs:

\vskip 0.5 cm

\begin{minipage}{15cm}
\begin{center}
\includegraphics[width=0.60\textwidth]{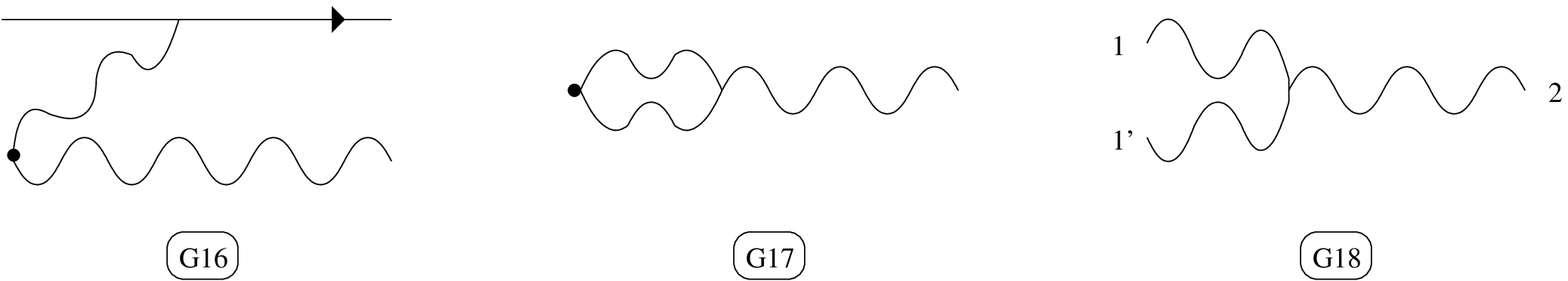}
\end{center}
\end{minipage}
\begin{center}
Figure 2. Additional graphs at order $g^4$.
\end{center}

Like in the pictures in \cite{mejan} we have omitted free matter propagators.
Point 1 is on the left and point 2 on the right of the graphs. The connection
carries the indices $\mu$ and $\nu$ there, respectively, while the connection
at 1' has index $\rho$. The lines are split only for convenience of drawing
--- the notation 1' in $G_{18}$ does not refer to a new point. It was
introduced in order to distinguish the two vector propagators joining the cubic
vertex from the left.

After the evaluation of Grassmann- and $SU(2)$ integrations we find
\begin{eqnarray}
G_{16} & = & (12) \, \bigr[ - \, \eta_{\mu \nu} / 4 \bigr] \, ,\\
G_{17} & = & \eta_{\mu \nu} / 2 \, , \\
G_{18} & = & i \, \bigr[ (\partial_1 - \partial_{1'})_\nu
\, \eta_{\mu \rho} / 4 \, - \, (\partial_1 - \partial_2)_\rho
\, \eta_{\mu \nu} / 4 \, + \, (\partial_{1'} - \partial_2)_\mu
\, \eta_{\nu \rho} / 4 \, \bigr] \, .
\end{eqnarray}
In the same way as graphs $G_3, \, G_4$ in \cite{mejan} occur together, we
may add $G_{16}$ and its mirror image $\tilde G_{16}$ into $G_{11}$ because
their combinatorics is equal:
\begin{eqnarray}
G_{11} \, + \, G_{16} \, + \, \tilde G_{16} & = & (12) \, \bigl[ \, - \,
\partial_{\nu 13} \partial_{\mu 23} \, + \, \eta_{\mu \nu} (\partial_{13} -
\partial_{23}). (\partial_{14} - \partial_{24}) / 4 \\
&& \phantom{(12) \, \bigl[ } \, + \, \eta_{\mu \nu}
(\square_{14} \, + \, \square_{24}) / 4 \, + \, \ldots \, \bigr] \nonumber
\end{eqnarray}
(The dots indicate omitted finite terms.)

The rest of the calculation proceeds along the same lines as before
(appropriately adapted to the new tensor component), i.e.~the reconstruction
of the Fourier transform of integrals with open indices from projections with
the total momentum $q$ and the $\eta$ symbol, which are built in
\emph{Mathematica} and evaluated by the \emph{Mincer} package \cite{mincer}.

\subsection{The Length 3 Spin 3 Mixing Problem}

As an illustration of what has been said before we re-examine the
mixing of the length three operators
\begin{equation}
\{s_1, \, s_2, \, s_3\} \, = \, Tr \bigl((D^{s_1}_z Z)(D^{s_2}_z
Z)(D^{s_3}_z Z)\bigr)
\end{equation}
at leading order in $N$. In particular, the spin three mixing
problem involves the operators
\begin{equation}
{\cal B} \, = \, \{ \, \{3,0,0\}, \, \{2,1,0\}, \, \{1,2,0\}, \,
\{1,1,1\} \, \} \, .
\end{equation}
The one-loop logarithms and the constant order $T_0$ of the tree-level
correlators are diagonalized by choosing the directions
\begin{eqnarray}
{\cal O} & = & \{1, \, 3, \, 3, \, 2\} \, , \\ {\cal K} & = & \{1,
\, -1, \, -1, \, -2 \} \, , \nonumber \\ {\cal V}_1 & = & \{2, \,
-9, \, -9, \, 24\} \, , \nonumber \\ {\cal V}_2 & = & \{0, \, 1,
\, -1, \, 0\} \nonumber
\end{eqnarray}
relative to the basis ${\cal B}$. Note that ${\cal V}_1, \, {\cal
V}_2$ have identical first anomalous dimension and therefore the
eigenspace may be spanned by any two independent directions. We have
split into an even and an odd part under reversal of the trace; as
a consequence ${\cal V}_2$ decouples from the other operators.
Renormalization in the $\overline{MS}$ scheme outlined above
yields the anomalous dimensions
\begin{eqnarray}
\gamma_{\cal O} & = &  0 \, , \\ \gamma_{\cal K} & = & g^2 \; \,
4 \; \, \, - \, g^4 \; \, 6 \, , \nonumber
\\  \gamma_{{\cal V}_1} & = &  g^2 \, \frac{15}{2} \, - \, g^4 \,
\frac{225}{16} \; , \nonumber \\ \gamma_{{\cal V}_2} & = & g^2 \,
\frac{15}{2} \, - \, g^4 \, \frac{225}{16} \; , \nonumber
\end{eqnarray}
up to terms of $O(g^6)$.
The individual ${\cal Z}_i$ are given by the anomalous dimensions
in the standard way. As explained above, the system does not entirely
determine the $C$ matrices, while we can fix $B$:
\begin{equation}
B \, = \,
\begin{pmatrix}
\phm 0 & \phm 0 & \phm 0 & \phm 0 \\ - \frac{1}{2} & \phm 0 & \phm
0 & \phm 0
\\ \phm \frac{3}{4} & - \frac{165}{28} & \phm 0 & \phm \alpha \\ \phm 0 &
\phm 0 & - 315 \, \alpha & \phm 0 \\
\end{pmatrix}
\end{equation}
The parameter $\alpha$ is not calculable from our system of
equations because the anomalous dimensions of ${\cal V}_1, \,
{\cal V}_2$ are degenerate. We may put it to zero bearing in mind that an
arbitrary remixing of the two operators is possible.

The anomalous dimensions and the entries of $B$ are independent of
whether we calculate the $(\eta_{z \bar z})^3$ terms as outlined in this
article or the $(x_z \, x_{\bar z})^3$ part of the correlators as
in \cite{mejan}, although now in $\overline{MS}$.\footnote{In \cite{mejan} we
deviated from the strict $\overline{MS}$ prescription by choosing the
Basis ${\cal B}$ in an $\epsilon$ dependent way, so that $T_1$ became diagonal
as well. This had the advantage of decoupling the protected operator
${\cal O}$.}

The operators ${\cal V}_1$ and ${\cal V}_2$ are conformal primaries of spin
three. The numerator of their renormalized two-point functions
should contain three powers of the inversion tensor $J_{z \bar z}
\, = \, \eta_{z \bar z} - 2 x_z x_{\bar z} / x^2$, and
correspondingly we find that the normalization of the $(x_z \,
x_{\bar z})^3$ terms differs by $-8$ from that of the $(\eta_{z
\bar z})^3$ part. The operator ${\cal K}$ is a first derivative of
the primary ${\cal K}_6$ \cite{mejan}. The normalizations of the
two terms in $\langle {\cal K} \, \bar {\cal K} \rangle$ are
indeed consistent with being derivatives of a common spin two
two-point function; similarly for the protected operator ${\cal O}
\, = 1/3 \, D^3_z \, \{0,0,0\}$. In conclusion, in this example
renormalization works in the same way for these two components of the
tensor structure. Conformal invariance is manifest.

\section{The Dilatation Operator and Renormalization}
\label{app:dila}

\subsection{Matrix Elements of the Dilatation Operator in Dimensional
Regularization}

Suppose there are linear operators $\tilde D_1, \, \tilde D_2$
\begin{eqnarray}
\tilde D_1 \, {\cal O}_i & = & \bigl( \frac{1}{\epsilon} \, D_{11
\, ij} \, + \, D_{10 \, ij} \bigr) {\cal O}_j \, , \\ \tilde D_2
\, {\cal O}_j & = & \bigl( \frac{1}{\epsilon^2} \, D_{22 \, ij} \,
+ \, \frac{1}{\epsilon} \, D_{21 \, ij} \bigr) {\cal O}_j \, ,
\end{eqnarray}
such that
\begin{eqnarray}
\langle {\cal O}_i \, \bar {\cal O}_j \rangle_{g^2} & = \,
\langle (\tilde D_1 {\cal O}_i) \, \bar {\cal O}_j \rangle_{g^0}
\, = & \bigl( \frac{1}{\epsilon} \, D_{11} \, T_0 \, + \, ( D_{10}
\, T_0 \, + \, D_{11} \, T_1 ) \bigr)_{ij} \, , \label{twog2} \\ \langle {\cal
O}_i \, \bar {\cal O}_j \rangle_{g^4} & = \, \langle (\tilde
D_2 {\cal O}_i) \, \bar {\cal O}_j \rangle_{g^0} \, = & \bigl(
\frac{1}{\epsilon^2} \, D_{22} \, T_0 \, + \, \frac{1}{\epsilon}
\, ( D_{21} \, T_0 \, + \, D_{22} \, T_1 ) \bigr)_{ij} \, . \label{twog4}
\end{eqnarray}
The eigenvectors of $D_{11}$ constitute the aforementioned diagonal
basis ${\cal O}_i$. In this frame
$D_{11} \, = - \, \Gamma_1$, by which token the pole part of $\tilde D_1$
is the negative of the \emph{one-loop dilatation operator}.

We will now consider the epsilon expansion of equation (\ref{starthere}) order
by order in $g^2$ up to $O(\epsilon)$. For the rest of this section we
assume the operators to be eigenvectors of $D_{11}$.

We may take $X(L,s)$ out of our system of equations: any set of
renormalization factors, that renders finite the bare correlators without
the $X(L,s)$ factor, remains a solution on multiplication by $X(L,s)$ because
the latter is not singular in $\epsilon$.

{}From the constant part at $g^0$ we immediately identify
$a_{0 \, i} \, = \, t_{0 \, ii}$. At $O(g^2)$ the epsilon expansion
yields simple logarithms, simple poles and a constant part.
{}From the first two
sets of terms and the diagonal of the third we learn
\begin{equation}
\gamma_{1 \, i} \, = \, z_{11 \, i} \, = \, - D_{11 \, ii} \, , \qquad
a_{1 \, i} \, = \, D_{10 \, ii} \, t_{0 \, ii}\, ,
\end{equation}
while the off-diagonal part of the constant term constrains $B$
but is not sufficient to fix it completely; hermiticity of the two-point
function on the l.h.s. of (\ref{twog2}) halves the number of
independent equations. (This places constraints on $D_{10}$. Similarly
$D_{21}$ is constrained by the hermiticity of the l.h.s. of (\ref{twog4}).)

At $O(g^4)$ there is a number of conditions to solve: the double pole and
the double logarithm in the epsilon expansion of (\ref{starthere}) yield two
equations implying that
\begin{equation}
z_{22 \, i} \, = \, D_{22 \, ii} \, = \, \frac{1}{2} \gamma_{1 \, i}^2
\end{equation}
while the $\log(x_{12}^2) / \epsilon$ terms give nothing new. The diagonals of
the simple logarithm and simple pole parts lead to
\begin{equation}
\gamma_{2 \, i} \, = \, z_{21 \, i} \, = \, - 2 \, ( D_{21 \, ii} \, - \,
D_{10 \, ii} \, D_{11 \, ii}) \label{gam2sol} \, .
\end{equation}
The r.h.s. of the last equation is actually the action of a combination of
$\tilde D_1, \, \tilde D_2$:
\begin{equation}
\frac{1}{\epsilon} \, ( D_{21 \, ii} \, - \, D_{10 \, ii} \, D_{11 \, ii}) \,
= \, \bigl((\tilde D_2 \, - \, \frac{1}{2} \, \tilde D_1^2 ) \, {\cal O}_i
\bigr)_i
\end{equation}
The off-diagonal entries of the simple logarithm part
depend on $B$ and those of the simple pole part on $B$ and $C_1$. The
matrix $C_1$ cannot yet be fixed uniquely, but we now have enough equations to
compute $B$.  The resulting matrix equation is the off-diagonal part of
\begin{equation}
B \, D_{11} \, - \, D_{11} \, B \, = \,  - 2 \, \bigl(D_{21} \, - \frac{1}{2}
(D_{11} \, D_{10} \, + \, D_{10} \, D_{11}) \, - \, \frac{1}{4} (D_{11} \,
 D_{10} \, - \, D_{10} \, D_{11}) \bigr) \label{Beq}\, .
\end{equation}
Remarkably, the last term in this expression does not contribute on the
diagonal, because $D_{11}$ is diagonal. Hence the matrix
\begin{equation}
D_2 \, = \,  - 2 \, \bigl(D_{21} \, - \frac{1}{2}
\bigl\{D_{11}, D_{10}\bigr\} \, - \, \frac{1}{4} \bigl[D_{11}, \,
D_{10} \bigr] \bigr) \label{defdel}
\end{equation}
has $\gamma_{2 \, i}$ on its diagonal and it determines $B$ through
(\ref{Beq}).

It was shown in \cite{merome} that the \emph{two-loop dilatation generator}
acts in precisely this way: suppose that the dilatation operator has an
expansion
\begin{equation}
\Delta \, = \, 1 \, + \, g^2 \, \Delta_1 \, + \, g^4 \, \Delta_2 \, +
\, \ldots \, .
\end{equation}
We want to solve the eigenvalue problem
\begin{equation}
\Delta \, \bigl( {\cal O} \, + \, g^2 \, B \, {\cal O} \, + \ldots \bigr)
 \, = \, \bigl(1 \, + \, g^2 \, \Gamma_1 \, + \, g^4 \, \Gamma_2 \, +
\ldots \bigr) \bigl( {\cal O} \, + \, g^2 \, B \, {\cal O} \, + \ldots
\bigr) \, .
\end{equation}
Here $\Gamma_1, \, \Gamma_2$ are diagonal matrices containing the anomalous
dimensions of the individual operators, and the lowest order re-mixing of the
operators is named $B$. The dilatation operator acts on the vector of operators
${\cal O}$ as a linear map
\begin{equation}
\Delta_1 \, {\cal O} \, = \, D_1 \, {\cal O} \, , \qquad \Delta_2 \, {\cal O}
\, = \, D_2 \, {\cal O} \, .
\end{equation}
Once again, we choose the basis for the operators to be the set of eigenvectors
of $D_1$, so that $\Delta_1 \, {\cal O} \, = \, D_1 \, {\cal O} \, = \,
\Gamma_1 \, {\cal O}$. The eigenvalue problem at order $g^4$ yields
\begin{equation}
D_2 \, = \, \Gamma_2 \, + \, (\Gamma_1 \, B \, - \, B \, \Gamma_1)
\end{equation}
exactly like $D_2$ from (\ref{defdel}). Note that the diagonal of $B$
remains undetermined --- it corresponds to trivial operator rescalings and may
be put to zero.

We have thus identified the matrix elements of the two-loop dilatation
operator from the renormalization procedure in dimensional regularization.
The next section addresses the construction of the dilatation operator itself.

\subsection{The one-loop dilatation operator}

In the planar limit the combinatorics for the two-point functions
$\langle {\cal X} \, \bar {\cal Y} \rangle$ has the following features:
\begin{itemize}
\item At tree-level, we find a cyclic sum over, say, site $1$ in ${\cal X}$
joining site $i$ in $\bar {\cal Y}$. All other lines are parallel.
\item At loop-level, the interaction is between adjacent sites. It can occur
at any site in each part of the tree-level configuration.
\end{itemize}
The $O(g^2)$ contribution to the correlator $\langle {\cal X} \,
\bar{\cal Y} \rangle$ originates from the ${\cal N} = 2$ supergraphs

\vskip 0.5 cm

\begin{minipage}{15cm}
\begin{center}
\includegraphics[width=0.80\textwidth]{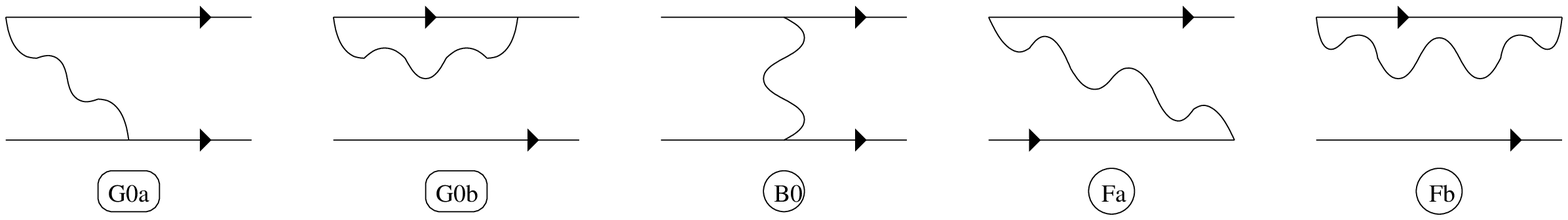}
\end{center}
\end{minipage}
\begin{center}
Figure 3. Graphs defining the one-loop dilatation operator.
\end{center}

\noindent where, of course, the underlying Feynman integral is the same in
$G_{0a}, \, G_{0b}$. It was called $G_0$ in \cite{mejan} and is one-loop
divergent. The ``BPS-like'' graph $B_0$ is finite. The third structure
$F$ simply has a free vector line; it involves no loop-integration.
The configurations $G_{0a}, \, G_{0b}$ occur with the gauge line emanating
from any of the four end-points; likewise, $F_a, \, F_b$ must be joined by
the opposite constellations.

It is natural to interpret the one-loop interaction as a sum over a two-site
``Hamiltonian'' shifting over all sites in ${\cal X}$, which is then
contracted on $\bar {\cal Y}$ much as in the tree-level correlator. The
combinatoric factors for the Feynman graphs can be found by looking at the
correlator
\begin{equation}
F_1(s_1,s_2,s_3,s_4) \, = \, \langle \, Tr(T^a \, D_z^{s_1} Z \, D_z^{s_2}
Z)(1) \; Tr(D_{\bar z}^{s_4} \bar Z \, D_{\bar z}^{s_3} \bar Z \, T^b)(2) \,
\rangle \,  \label{excise1}
\end{equation}
at leading order in $N$ (i.e.~$N^2$), which is in a manner of speaking the
one-loop interaction excised from the full correlator $\langle {\cal X} \,
\bar {\cal Y} \rangle$. We find a -2 for the ``disconnected parts''
$G_{0b}, F_b$ and a 1 otherwise. The disconnected diagrams
can be attributed to the two-site interaction to their left or to their right,
so that we scale by 1/2 in order to avoid over-counting.

If the interaction connects sites $i,i+1$ in ${\cal X}$ to $j,j+1$ in
${\cal Y}$, then the other fields in the operators are joined by parallel free
lines
\begin{equation}
\Pi(s_1,s_2) \, = \, \partial_{z\, 1}^{s_1} \partial_{\bar z \, 2}^{s_2} \,
\Pi_{12} \, = \, - \delta_{s_1,s_2} \, \frac{\eta_{z \bar z}^{s_2} \, 2^{s_2}
\, s_2! \,  \sum_{k=1}^{s_2} (k-\epsilon)}{4 \pi^2 (x_{12}^2)^{(s2+1-
\epsilon)}} \, + \, \ldots \, ,
\end{equation}
where the omitted terms contain $x_z$ or $x_{\bar z}$. The key observation is
that the $X(L,s)$ term in the complete correlator can only exists when all free
lines have the same spin at both ends \cite{oneloop}. Coupling between sites
with different spin is only possible where the interaction is; since we want no
$x_{12}$ with free indices the interaction can at most ``transfer'' a
derivative from one of the two sites to the other. In particular, it must
conserve the total spin.

Let us normalize by the inverse of the tree-level. This will simply remove
all the free lines and scale down $F_1$
\begin{equation}
{\cal H}_i^{(0)} \, = \, \hat F_1(s_i,s_{i+1},s_j,s_{j+1}) \frac{1}{\Pi(s_j,s_j) \,
\Pi(s_{j+1},s_{j+1})} \label{defhi}\, ,
\end{equation}
where $\hat F_1$ is $F_1$ with the over-counting corrected and the
group factor $N^2 \delta^{ab}$ stripped off.

Whithout any derivatives, the graphs $G_{0a}, \, G_{0b}, \, F_a, \, F_b$
are absent while $B_0 \, = \, O(\epsilon)$, whence $\hat F_1(0,0,0,0) \,
\rightarrow \, 0$. When the total spin is not zero, ${\cal H}_i^{(0)}(s)$ is
conveniently given as a matrix: \\
At \textbf{spin 1} we can have
\begin{equation}
\{s_i,s_{i+1}\}, \, \{s_j,s_{j+1}\} \, \in \, \{ \{1,0\}, \, \{0,1\} \}
\end{equation}
and our set of graphs produces
\begin{equation}
{\cal H}_i^{(0)}(1) \, = \, - \frac{1}{\epsilon} \,
\left( \begin{array}{ll} \phantom{-}1 & -1 \\ -1 & \phantom{-}1 \end{array}
\right) \, .
\end{equation}
At \textbf{spin 2} we have the basis\footnote{The normalization of the basis
elements reflects the fact that several derivatives at the same site are
indistinguishable.}
\begin{equation}
\{s_i,s_{i+1}\}, \, \{s_j,s_{j+1}\} \, \in \, \{ \frac{1}{2} \{2,0\}, \,
\{1,1\}, \, \frac{1}{2} \{0,2\} \}
\end{equation}
and the rules for transferring derivatives are
\begin{equation}
{\cal H}_i^{(0)}(2) \, = \, - \frac{1}{\epsilon}
\left( \begin{array}{lll}
                   \phantom{-}\frac{3}{2} & -1 & -\frac{1}{2} \\
                   -1 & \phantom{-}2 & -1 \\
                   -\frac{1}{2} & -1 & \phantom{-}\frac{3}{2}
                  \end{array}
                  \right) \, + \,  \left(
                  \begin{array}{lll}
                   \phantom{-}\frac{1}{2} & \phantom{-}0 & \phantom{-}0 \\
                   -\frac{1}{2} & \phantom{-}0 & -\frac{1}{2} \\
                   \phantom{-}0 & \phantom{-}0 & \phantom{-}\frac{1}{2}
                  \end{array}
                  \right) \, .
\end{equation}
At \textbf{spin 3} the basis elements are
\begin{equation}
\{s_i,s_{i+1}\}, \, \{s_j,s_{j+1}\} \, \in \, \{ \frac{1}{6} \{3,0\}, \,
\frac{1}{2} \{2,1\}, \, \frac{1}{2} \{1,2\}, \, \frac{1}{6} \{0,3\} \}
\end{equation}
while the derivatives may be transferred according to
\begin{equation}
{\cal H}_i^{(0)}(3) \, = \, - \frac{1}{\epsilon} \left(
                  \begin{array}{llll}
                \phantom{-}\frac{11}{6} & -1 & -\frac{1}{2} & -\frac{1}{3} \\
                -1 & \phantom{-}\frac{5}{2} & -1 & -\frac{1}{2} \\
                -\frac{1}{2} & -1 & \phantom{-}\frac{5}{2} & -1 \\
                -\frac{1}{3} & -\frac{1}{2} & -1 & \phantom{-}\frac{11}{6}
                  \end{array}
                  \right) \, + \, \left(
                  \begin{array}{llll}
                   \phantom{-}1 & \phantom{-}0 & \phantom{-}0 & \phantom{-}0 \\
         -\frac{2}{3} & \phantom{-}\frac{1}{2} & -\frac{1}{2} & -\frac{1}{3} \\
         -\frac{1}{3} & -\frac{1}{2} & \phantom{-}\frac{1}{2} & -\frac{2}{3} \\
                   \phantom{-}0 & \phantom{-}0 & \phantom{-}0 & \phantom{-}1
                  \end{array}
                  \right) \, .
\end{equation}
The pole part of these rules accurately reproduces the result of
\cite{oneloop}:
The diagonal entries are $h(s_i) + h(s_{i+1})$ where $h(n)$ are the harmonic
numbers, and the off-diagonal entries are $-1/d$ where $d$ counts the number
of transferred derivatives. The finite part could doubtlessly also be fitted:
we observed that the contribution from $B_0$ apparently always equals that of
$F_a,F_b$ which is trivial to compute. Graphs $G_{0a},G_{0b}$ contain
only a one-loop integral, so that a result can be obtained in closed form.
On the contrary, at the two-loop level this is not easy due to the complexity
of the integrals. Consequently, we limit the scope of this work to the first
few cases obtained by direct calculation.

The one-loop dilatation operator is defined as
\begin{equation}
\tilde D_1 \, = \, \sum_{i=1}^l \, {\cal H}_i \, ,
\end{equation}
i.e.~the ``Hamiltonian'' runs over all sites in an operator ${\cal X}$, mapping
it to a sum of terms with a new distribution of the derivatives over the sites
in the chain. By construction,
\begin{equation}
\langle (\tilde D_1 \, {\cal X}) \, \bar {\cal Y} \rangle_{g^0} \, = \,
\langle {\cal X} \, \bar {\cal Y} \rangle_{g^2} \, .
\end{equation}

We conclude the section with two remarks: first, the definition of
${\cal H}_i^{(0)}$ in (\ref{defhi}) is necessarily asymmetric because we have
normalized from the right. Correspondingly, the constant parts of the transfer
rules are not
symmetric matrices. On the other hand, the pole part is symmetric, because
in terms of complete two-point functions the matrices $T_0$ and $\Gamma_1$ must
be simultaneously diagonalizable.
Second, it should be stressed that the $X(L,s)$ terms are by far better suited
to the construction of the interaction Hamiltonian ${\cal H}_i$ than for
example the terms with no traces considered in \cite{mejan}: those allow
non-vanishing
free lines between $D^{s_1}_z Z(1)$ and $D^{s_2}_{\bar z} \bar Z(2)$ for
unequal spins $s_1 \, \neq \, s_2$, and the interaction need not conserve the
total spin either. While the pole part of the one-loop dilatation operator
is correctly obtained in this picture, we found it problematic to
consistently subtract out disconnected parts at two loops.

\subsection{The two-loop dilatation operator}

In analogy to (\ref{excise1}) we try to read off the operator $\tilde D_2$
from the $O(g^4)$ contribution to
\begin{equation}
F_2(s_1,s_2,s_3,s_4,s_5,s_6) \, = \, \langle \, Tr(T^a \, D_z^{s_1} Z \,
D_z^{s_2} Z \, D_z^{s_3} Z)(1) \; Tr(D_{\bar z}^{s_6} \bar Z \,
D_{\bar z}^{s_5} \bar Z \, D_{\bar z}^{s_4} \bar Z \, T^b)(2) \,
\rangle \, .
\end{equation}
In doing so we should remember that matrix elements of the two-loop dilatation
operator were defined by several terms, most prominently $\gamma_2$ came about
as a matrix element of the combination $\tilde D_2 - \tilde D_1^2 / 2$, see
equation (\ref{gam2sol}) and the comment after it. We fall on the
renormalization scheme of \cite{BKS}: the two-loop effective vertex has to
be corrected by subtracting the square of the one-loop vertex. Explicitly,
we take out
\begin{eqnarray}
\frac{1}{2} \, \tilde D_1^2 & = & \frac{1}{2} \, \sum_i {\cal H}^{(0)}_i \, \sum_j {\cal H}^{(0)}_j \label{allsquared} \\
& = & \sum_{i+1<j} {\cal H}^{(0)}_i \, {\cal H}^{(0)}_j \, + \, \frac{1}{2} \, \sum_i \Bigl( \frac{1}{2}
\, ({\cal H}^{(0)}_i)^2 \, + \, {\cal H}^{(0)}_i {\cal H}^{(0)}_{i+1} \, + \, {\cal H}^{(0)}_{i+1} {\cal H}^{(0)}_i \, + \, \frac{1}{2} \,
({\cal H}^{(0)}_{i+1})^2 \Bigr) \nonumber \, .
\end{eqnarray}
(The derivation of the dilatation operator presented here is ``asymptotic''
in that it assumes the existence of disconnected pieces.) The first term in
the last formula corresponds to the situation where the two one-loop
Hamiltonians do not overlap, thus all terms are disconnected. If both pairs
$\{i,i+1\}, \, \{j,j+1\}$ are outside our ``window'' $F_2$, they will simply
cancel disconnected parts that we do not see in the excised part. Likewise,
if only one of ${\cal H}^{(0)}_i, \, {\cal H}^{(0)}_j$ touches the excised part, we would see an order
$g^2$ contribution, which we need not consider. Thus the cases of interest
are (we put the left of $F_2$ at position $i$)
\begin{equation}
(i) \quad {\cal H}^{(0)}_{i} {\cal H}^{(0)}_{i+2} \, ,
\qquad (ii) \quad {\cal H}^{(0)}_{i-1} {\cal H}^{(0)}_{i+1} \, ,
\qquad (iii)
\quad {\cal H}^{(0)}_{i-1} {\cal H}^{(0)}_{i+2}\, ,
\label{d1squared}
\end{equation}
whose relevant $g^4$ diagrams may be directly subtracted from the set of
graphs in $F_2$. The second term in (\ref{allsquared}) is unfortunately not
amenable to this treatment: by way of example we do not have a diagram that
identically equals two consecutive contributions of $G_{0a}$.

Our strategy thus starts by setting up an operator $J_i$ from the $g^4$
graphs in $F_2$ with the subtraction of disconnected parts described in the
last paragraph, whereas the overlapping part of $(\tilde D_1)^2/2$ will be
dealt with later on. To avoid over-counting we have to rescale
contributions with free lines: in complete analogy to the one-loop case we
scale down by a factor 1/2 such graphs, that connect two matter lines but leave
the right or left line free. Note that no re-scaling is needed when the free
line is the central one; this situation is particular to exactly one position
of the Hamiltonian. Configurations with
two free lines can be arbitrarily shifted between the three positions within
the Hamiltonian because the dilatation operator will involve a sum over
positions. In order to compensate over-counting we choose to scale by 1/4
if the interaction is concentrated on one of the outer lines, and
by 1/2 if it is on the central line. We define
\begin{equation}
J_i \, = \, \hat F_2(s_{i},s_{i+1},s_{i+2},s_{j},s_{j+1},
s_{j+2}) \frac{1}{\Pi(s_{j},s_{j}) \, \Pi(s_{j+1},s_{j+1}) \,
\Pi(s_{j+2},s_{j+2})}\, ,
\label{defji}
\end{equation}
with $\hat F_2$ being $F_2$ after the appropriate modification of the set of
graphs and once again after omission of the group factor $N^3 \delta^{ab}$.

The connected part in (\ref{allsquared}) can be derived from the transfer
rules for derivatives given in the last section. Recall that according to
equation (\ref{defdel}) the matrix elements of the two-loop dilatation operator
also contain the term $-1/4 (D_{11} D_{10} - D_{10} D_{11})$, when the
dilatation operator is made to reproduce the $O(g^2)$ remixing $B$.
By splitting the one-loop transfer rules into a pole part
${\cal H}^{(0)}_{1 \, i}$ and a
constant piece ${\cal H}^{(0)}_{0 \, i}$, we can construct this term as an
operator in much
the same way as the connected part of $\tilde D_1^2$. Note that
$-1/4 [ {\cal H}^{(0)}_{1,
\, i}, {\cal H}^{(0)}_{0, \, j} ]$ has no disconnected part since
${\cal H}^{(0)}_1$ and ${\cal H}^{(0)}_0$ commute
when they do not overlap.

Finally, the full two-loop dilatation operator takes the form
\begin{equation}
D_2 \, = \, \sum_{i=1}^l \, {\cal H}^{(2)}_i|_{\epsilon^{-1}}\, ,
\end{equation}
with the two-loop Hamiltonian
\begin{eqnarray}
{\cal H}^{(2)}_i & = & J_i \label{twohamilton} \\
& - & \frac{1}{2} \, \Bigl( \frac{1}{2} \, ({\cal H}^{(0)}_i)^2 \,
+ \, {\cal H}^{(0)}_i {\cal H}^{(0)}_{i+1} \, +
\, {\cal H}^{(0)}_{i+1} {\cal H}^{(0)}_i \, + \, \frac{1}{2} \,
({\cal H}^{(0)}_{i+1})^2 \Bigr) \nonumber \\ & - &
\frac{1}{4} \Bigl( \frac{1}{2} \,
{\cal H}^{(0)}_{1 \, i} {\cal H}^{(0)}_{0 \, i} \, + \,
{\cal H}^{(0)}_{1 \, i} {\cal H}^{(0)}_{0 \,i+1} \,
+ \, {\cal H}^{(0)}_{1 \, i+1} {\cal H}^{(0)}_{0 \, i} \, +
\, \frac{1}{2} \, {\cal H}^{(0)}_{1 \, i+1} {\cal H}^{(0)}_{0 \, i+1} \,
- ({\cal H}^{(0)}_1 \leftrightarrow {\cal H}^{(0)}_0)
\Bigr) \nonumber \, .
\end{eqnarray}
The Hamiltonian ${\cal H}^{(2)}_i$ has in fact a non-vanishing $1/\epsilon^2$
part, but the second order poles are distributed over the matrices
in such a way that they drop in the sum over all positions. The transfer
rules below and in the main text describe the $1/\epsilon$ part.

In this appendix we give the transfer rules corresponding to the full
Hamiltonian including the terms in the last line of (\ref{twohamilton}).
These come from the commutator
$$
-1/4 \, [ {\cal H}^{(0)}_{1 \, i}, {\cal H}^{(0)}_{0 \, j} ].
$$
Note that this term is anti-hermitian,
so that the transfer rules cannot be transformed
into symmetric matrices. In the main text we omit the commutator term, since
in the context of the Bethe ansatz it is preferable to have a hermitian
Hamiltonian. In any case, the exact resolution of the mixing is not easy to
obtain in the Bethe ansatz picture which projects out the descendants.
Surprisingly, the formulae below do apply to the length 3 spin 3 mixing
problem although they were derived for longer chains for which
disconnected pieces have to be subtracted.
\newline
The explicit bases and two-loop transfer rules up to spin 3 are:
\newline
\begin{minipage}{15cm}
\noindent \textbf{Spin 1} \\
basis: $\{\{1,0,0\},\, \{0,1,0\}, \, \{0,0,1\}\}$
\begin{equation}
{\cal H}^{(2)}_i(1) \, = \, \left( \begin{array}{lll}
                  -\frac{3}{4} & \phm 1 & -\frac{1}{2} \\
                  \phm 1 & -\frac{3}{2} & \phm 1 \\
                  -\frac{1}{2} & \phm 1 & -\frac{3}{4}
                 \end{array}
                 \right)
\nonumber\end{equation}
\end{minipage}
\newline
\begin{minipage}{15cm}
\noindent \textbf{Spin 2} \\
basis: $\{\frac{1}{2}\{2,0,0\},\, \{1,1,0\}, \, \{1,0,1\}, \, \frac{1}{2}
\{0,2,0\}, \, \{0,1,1\}, \, \frac{1}{2} \{0,0,2\}\}$
\begin{equation}
{\cal H}^{(2)}_i(2) \, = \, \left(
   \begin{array}{llllll}
    -\frac{19}{32} & \phm \frac{17}{16} & -\frac{1}{2} & \phm \frac{1}{2} &
-\frac{1}{4} & -\frac{1}{16} \\
    \phm \frac{21}{16} & -\frac{9}{4} & \phm 1 & \phm \frac{29}{16} & \phm 0 &
-\frac{1}{8} \\
    -\frac{3}{4} & \phm 1 & -\frac{3}{2} & -\frac{1}{2} & \phm 1 &
-\frac{3}{4} \\
    \phm \frac{1}{4} & \phm \frac{11}{16} & \phm 0 & -\frac{67}{16} & \phm
\frac{11}{16} & \phm \frac{1}{4} \\
    -\frac{1}{8} & \phm 0 & \phm 1 & \phm \frac{29}{16} & -\frac{9}{4} & \phm
\frac{21}{16} \\
    -\frac{1}{16} & -\frac{1}{4} & -\frac{1}{2} & \phm \frac{1}{2} &
\phm \frac{17}{16} & -\frac{19}{32}
   \end{array}
   \right)
\nonumber\end{equation}
\end{minipage}
\newline
\begin{minipage}{15cm}
\noindent \textbf{Spin 3} \\
basis:
$\{\frac{1}{6} \{3,0,0\}, \, \frac{1}{2} \{2,1,0\}, \, \frac{1}{2} \{2,0,1\},
\, \frac{1}{2} \{1,2,0\}, \, \{1,1,1\}, \frac{1}{2} \{1,0,2\},$ \newline
$~~~~~~~~~~~\frac{1}{6} \{0,3,0\}, \, \frac{1}{2} \{0,2,1\}, \,
\frac{1}{2} \{0,1,2\}, \, \frac{1}{6} \{0,0,3\}\}$
\begin{equation}
{\cal H}^{(2)}_i(3) \, = \, \left(
   \begin{array}{llllllllll}
    \phm \frac{25}{288} & \phm \frac{11}{18} & -\frac{1}{2} & \phm
\frac{77}{144} & -\frac{1}{4} & -\frac{1}{16} & \phm \frac{71}{216} &
-\frac{1}{6} & -\frac{1}{24} & -\frac{1}{54} \\
   \phm \frac{43}{72} & -\frac{63}{32} & \phm 1 & \phm \frac{3}{2} & \phm 0 &
-\frac{1}{8} & \phm \frac{149}{144} & \phm 0 & -\frac{1}{16} & \phm
\frac{1}{72} \\
    -\frac{5}{6} & \phm 1 & -\frac{43}{32} & -\frac{1}{2} & \phm \frac{17}{16}
 & -\frac{3}{4} & -\frac{1}{3} & \phm \frac{1}{2} & -\frac{3}{8} &
-\frac{7}{48} \\
  \phm \frac{91}{144} & \phm \frac{5}{4} & \phm 0 & -\frac{81}{16} &
\phm \frac{11}{16} & \phm \frac{1}{4} & \phm \frac{191}{72} & \phm 0 & \phm 0 &
 -\frac{1}{18} \\
    -\frac{7}{24} & \phm 0 & \phm \frac{21}{16} & \phm \frac{29}{16} & -3 &
\phm \frac{21}{16} & -\frac{1}{6} & \phm \frac{29}{16} & \phm 0 &
-\frac{7}{24} \\
    -\frac{7}{48} & -\frac{3}{8} & -\frac{3}{4} & \phm \frac{1}{2} &
\phm \frac{17}{16} & -\frac{43}{32} & -\frac{1}{3} & -\frac{1}{2} & \phm 1 &
-\frac{5}{6} \\
    -\frac{1}{216} & \phm \frac{19}{144} & \phm 0 & \phm \frac{19}{18} &
\phm 0 & \phm 0 & -\frac{1031}{144} & \phm \frac{19}{18} & \phm \frac{19}{144}
& -\frac{1}{216} \\
    -\frac{1}{18} & \phm 0 & \phm \frac{1}{4} & \phm 0 & \phm \frac{11}{16} &
\phm 0 & \phm \frac{191}{72} & -\frac{81}{16} & \phm \frac{5}{4} &
\phm \frac{91}{144} \\
    \phm \frac{1}{72} & -\frac{1}{16} & -\frac{1}{8} & \phm 0 & \phm 0 & \phm 1
& \phm \frac{149}{144} & \phm \frac{3}{2} & -\frac{63}{32} & \phm
\frac{43}{72} \\
    -\frac{1}{54} & -\frac{1}{24} & -\frac{1}{16} & -\frac{1}{6} &
      -\frac{1}{4} & -\frac{1}{2} & \phm \frac{71}{216} & \phm \frac{77}{144} &
      \phm \frac{11}{18} & \phm \frac{25}{288}
   \end{array}
   \right)
\nonumber\end{equation}
\end{minipage}

\section{Explicit Solution for Twist-Two}
\label{app:hahn}

The one-loop Bethe equation
\eqref{1loopbethe} may be recast as a second-order difference
equation for the Baxter-$Q$ function $Q_s(u)$
\begin{equation}\label{Baxtereq}
T_s(u)\,Q_s(u)=(u+\frac{i}{2})^L\,Q_s(u+i)+(u-\frac{i}{2})^L\,Q_s(u-i)\, ,
\end{equation}
where $Q_s(u)$ is a polynomial of degree $s$ in the variable $u$,
whose algebraic roots are the Bethe roots $\{u_k\}$,
\begin{equation}\label{Baxterdef}
Q_s(u)=C_s\,\prod_{k=1}^{s}\, (u-u_k)\, ,
\end{equation}
i.e.~the solutions of \eqref{1loopbethe}, and
$C_s$ is an, for our purposes, irrelevant normalization constant.
For twist $L=2$ the excitation number $s$ has to be even,
and the Baxter equation \eqref{Baxtereq} is exactly solvable
in terms of a hypergeometric function
\begin{equation}\label{hahn}
Q_s(u)={}_3F_2[-s,s+1,\frac{1}{2}-i u;1,1;1]
\quad {\rm with} \quad
T_s(u)=2\,u^2-s^2-s-\frac{1}{2}\, .
\end{equation}
The hypergeometric series terminates if $s$ is an even natural number,
and therefore generates the explicit polynomial solution of the
twist-two Baxter equation\footnote{
The details of the solution \eqref{hahn} were worked out by
Virginia Dippel (unpublished) by adapting the method of
\cite{korchemsky} to the present case. The polynomials
\eqref{hahn} belong to the family of so-called
{\it Hahn polynomials} \cite{korchemsky}.
}.
The roots are all real and their distribution is even,
i.e.~the $Q_s(u)$ in \eqref{hahn} are actually polynomials in $u^2$.
Therefore the cyclicity constraint in \eqref{1loopmomeng} is
automatically satisfied. The energy is found from
\eqref{1loopmomeng},\eqref{Baxterdef} to be
\begin{equation}\label{hahneng}
E_s=2\,i\,\frac{d}{du}\,\left[\log Q_s(u+\frac{i}{2})\right]_{u=0}
=4\,\Big(\psi(s+1)-\psi(1)\Big)
=4\,h(s)\, .
\end{equation}
Here $h(s)=\sum_{j=1}^{s} 1/j$ are the harmonic numbers,
which may also be expressed through the logarithmic derivative
of the gamma function $\psi(s)=d/ds \log \Gamma(s)$.
In practice, the roots are easily found with a root finder.
E.g.~with \texttt{Mathematica} one may define
\begin{tabbing}
\verb"Hahn[s_, u_] := "
\\
\verb"Expand[HypergeometricPFQ[{-s, s + 1, 1/2 - I u}, {1, 1}, 1]]"
\end{tabbing}
and generate a table of all Bethe roots up to spin $s$, with an accuracy
of $k$ digits,
\begin{tabbing}
\verb"utable[s_] := Table[Flatten[NSolve[Hahn[2 t, u] == 0, u, k]], {t, 1, s/2}]"
\end{tabbing}
This is suitable, without further refinements, for finding the Bethe
roots up to spin $s\sim 70$ with an accuracy, if desired, of hundreds
of digits.

\section{Fourier Transforms}
\label{app:fourier}

\subsection{The Gauge Theory Ansatz}

In this appendix we find the Fourier Transform of the fourth term on
the r.h.s.~of (\ref{2loopintegral2}), in which is the the density
$\sigma(u')$ is integrated against the kernel
\begin{equation}
K(u,u') \, = \, i \, \partial_u \, \log \left(\frac{1 - g^2 / 2\,
x^+(u)\, x^-(u')}{1 - g^2 / 2\, x^-(u)\, x^+(u')} \right)^2\, .
\end{equation}
The definitions used in the last formula are
\begin{equation}
u \, = \, x(u) \, + \, \frac{g^2}{2 \, x(u)} \, , \qquad x(u) = \frac{u}{2}
\Bigl(1 \, + \, \sqrt{ 1 - \frac{2 g^2}{u^2}} \Bigr) \label{xdef}  \, ,
\end{equation}
\begin{equation}
u^\pm \, = \, u \, \pm \, \frac{i}{2} \, , \qquad x^\pm(u) \, = \, x(u^\pm)
\, .
\end{equation}
The branch cut of the square root is defined by the principal branch of the
logarithm. In the following we parametrize by
\begin{equation}
\tilde u^+ \, = \, \frac{1}{2} \, - \, i \, u \, = \, - \, i \, u^+ \, ,
\qquad \tilde u^- \, = \, \frac{1}{2} \, + \, i \, u \, = \, i \, u^- \, ,
\end{equation}
which obey the relation
\begin{equation}
\sqrt{ (\tilde u^\pm)^2 } \, = \, \tilde u^\pm \label{absoroot}
\end{equation}
because both $\tilde u^+, \, \tilde u^-$ have positive real part. Further, let
\begin{equation}
y(u) \, = \, \sqrt{ 1 + \frac{2 g^2 \lambda^2}{u^2} } \,
\end{equation}
and
\begin{equation}
K_0^\pm(u) \, = \, \frac{1}{\tilde u^\pm \, y(\tilde u^\pm)} \, , \qquad
K_1^\pm(u) \, = \, \frac{1}{\sqrt{2} g \lambda} \Biggl( 1 \, - \,
\frac{1}{y(\tilde u^\pm)} \Biggr) \, .
\end{equation}
Since we are, in this paper,
exclusively interested in \emph{symmetric} densities, we will
consider a $u' \, \leftrightarrow \, - \, u'$ symmetrized version of the
kernel. Our principal equation is
\begin{eqnarray}
&& - \, i \, g^2 \, \int_0^1 \, d\lambda \, \lambda \, \Bigl[ \; \phantom{+} \;
\partial_u \, \Bigl( K_0^+(u) \, - \, K_0^-(u) \Bigr) \, \Bigl( K_0^+(u') \,
+ \, K_0^-(u') \Bigr) \nonumber \\
&& \phantom{ i \, g^2 \, \int_0^1 \, d\lambda \, \lambda \, \Bigl[ }
+ \, \partial_u \, \Bigl(K_1^+(u) \, - \, K_1^-(u) \Bigr) \, \Bigl( K_1^+(u')
\, + \, K_1^-(u') \Bigr) \, \Bigr] \nonumber \\
&& = \, \frac{i}{2} \, \partial_u \, \log \left( \frac{ \bigl( 1 \, - \,
g^2 \, / \, 2 \,
x^+(u) \, x^-(u') \Bigr) \bigl( 1 \, + \, g^2 \, / \, 2 \, x^+(u) \,
x^+(u') \bigr) }{ \bigl( 1 \, - \, g^2 \, / \, 2 \, x^-(u) \,
x^+(u') \bigr) \bigl( 1 \, + \, g^2 \, / \, 2 \, x^-(u) \,
x^-(u') \bigr)} \right)^2 \, .
\end{eqnarray}
To prove this, we first do the parameter integrals on the left hand side:
\begin{eqnarray}
\int \, d\lambda \, \frac{\lambda}{y(\tilde u) \, y(\tilde u')} & = &
\frac{\tilde u \, \tilde u'}{2 \, g^2}
\log(\tilde u \, y(\tilde u) \, + \, \tilde u' \, y(\tilde u')) \, , \\
\int \, d\lambda \, \frac{1}{\lambda \, y(\tilde u) \, y(\tilde u')} & = &
\log (\lambda)
-\log(y(\tilde u) \, + \, y(\tilde u')) \, , \nonumber \\
\int \, d\lambda \, \frac{1}{\lambda y(\tilde u)} & = & \log(\lambda)
-\log(1 \, + \, y(\tilde u)) \, . \nonumber
\end{eqnarray}
Here we rely on (\ref{absoroot}) to simplify. Next, we change back to
the original variables $u^\pm$. We express the roots by $u,\, x(u)$ using
the second relation in (\ref{xdef}) in the form
\begin{equation}
\sqrt{ 1 - \frac{2 g^2}{u^2}} \, = \, \frac{ 2 \, x(u)}{u} \, - \, 1
\end{equation}
and finally eliminate $u$ in favor of $x(u), \, g^2$ by the first relation in
(\ref{xdef}). In a last step we collect all terms into one logarithm and
factor the argument. As long as $g$ is small this will not shift the
logarithm by some multiple of $\pi$; one may check that the Fourier
transform  below commutes with the Taylor expansion in $g$.

Next, we observe
\begin{equation}
K_j^\pm(u) \, = \, \int_0^\infty \, dt \, e^{\pm i \, u \, t} \, e^{-t/2} \,
J_j( \sqrt{2} g \lambda \, t) \, , \qquad j \, = \, 0,1 \label{ftobserve}
\end{equation}
and hence
\begin{eqnarray}
K_j^+(u) \, + \, K_j^-(u) \phantom{)} & = &  \int_{-\infty}^\infty \, dt \,
e^{i \, u \, t} \, e^{-|t|/2} \, J_j( \sqrt{2} g \lambda \, |t|) \, , \\
- \, i \, \partial_u \, (K_j^+(u) \, - \, K_j^-(u)) & = &
\int_{-\infty}^\infty \,
dt \, e^{i \, u \, t} \, |t| \, e^{-|t|/2} \, J_j( \sqrt{2} g \lambda \, |t|)
\nonumber \, .
\end{eqnarray}
Summing up, we have shown that
\begin{eqnarray}
&& \frac{i}{2} \, \partial_u \, \log \left( \frac{ \bigl( 1 \, -
\, g^2 \, / \, 2 \,
x^+(u) \, x^-(u') \bigr) \bigl( 1 \, + \, g^2 \, / \, 2 \, x^+(u) \,
x^+(u') \bigr) }{ \bigl( 1 \, - \, g^2 \, / \, 2 \, x^-(u) \,
x^+(u') \bigr) \bigl( 1 \, + \, g^2 \, / \, 2 \, x^-(u) \,
x^-(u') \bigr)} \right)^2 \label{showgauge} \\
& = &  g^2 \int_{-\infty}^\infty dt \, e^{i \, u \, t} \,
\int_{-\infty}^\infty dt' \, e^{i \, u' \, t'} \;
|t| \, e^{- (|t| \, + \, |t'|)/2} \; \hat K(\sqrt{2} g \, |t|, \, \sqrt{2} g \,
|t'|) \, , \nonumber
\end{eqnarray}
where
\begin{eqnarray}
\hat K(t, \, t') & = & \int_0^1 d\lambda \, \lambda \, \Bigl[ \,
J_0(\lambda \, t)  \, J_0(\lambda \, t') \, + \, J_1(
\lambda \, t)  \, J_1(\lambda \, t') \, \Bigr] \nonumber \\
& = & \frac{ J_1(t) \, J_0(t') \, -  \, J_0(t) \, J_1(t')}{t \, - \, t'} \, .
\end{eqnarray}
We conclude that for symmetric $\sigma(u')$
\begin{eqnarray}
&& e^{-|t|/2} \, \int_{-\infty}^\infty du \, e^{- i \, t \, u} \;
\int_{-\infty}^{\infty} du' \, K(u, \, u') \; \sigma(u') \label{finalform} \\
& = & 2 \, \pi \, g^2 \, |t| \, e^{-|t|} \, \int_{-\infty}^\infty dt' \,
\hat K(\sqrt{2} g \, |t|, \, \sqrt{2} g \, |t'|) \, \left[ \, e^{-|t'|/2}
\int_{-\infty}^\infty du' \, e^{ i \, u' \, t'} \, \sigma(u') \, \right]
\nonumber \\  & = & 2 \, \pi \, g^2 \, |t| \, e^{-|t|} \,
\int_{-\infty}^\infty dt' \,
\hat K(\sqrt{2} g \, |t|, \, \sqrt{2} g \, |t'|) \; \hat \sigma(t') \, ,
\nonumber
\end{eqnarray}
where we have used \eqref{fourier}.
Now, $\hat \sigma(t')$ is an even function if $\sigma(u')$ is.
We may thus reduce to the positive half axis, which yields the final form
of the last term in (\ref{prefredholm}).

Similar to formula (\ref{ftobserve}) one has
%
\begin{eqnarray}
&& \int_0^\infty \, dt \, e^{\pm i \, u \, t} \, e^{-t/2} \,
\frac{J_j(\sqrt{2} g \, t)}{\sqrt{2} g \, t} \\
& = & \frac{(\sqrt{2} \, g)^{j-1}}{j} \, \left( \tilde u^\pm \left(1 \, +
\, \sqrt{1 \, + \, 2 \, g^2 \, / \, (\tilde u^\pm)^2} \right) \, \right)^{- \,
j} \, , \quad j \, \geq \, 1 \, . \nonumber
\end{eqnarray}
From this one can easily derive
the following pretty result for the Fourier transforms
$\hat q_r(t)$ of the eigenvalues $q_r(u)$ of the commuting operators
of the integrable magnet. The expression \cite{BDS}
\begin{equation}
q_r(u)=\frac{1}{r-1}\,
\left(\frac{i}{x^+(u)^{r-1}}-\frac{i}{x^-(u)^{r-1}}\right)\, ,
\end{equation}
turns into\footnote{
These expressions were first obtained by Didina Serban
(2005, unpublished).}
\begin{equation}
\hat q_r(t)=\int_{-\infty}^{\infty}du\,e^{-i\,t\,u}\,q_r(u)=
4 \, \pi \, \left(\frac{\sqrt{2}}{i \, g} \right)^{r-2} \, e^{-|t|/2}
\, \frac{J_{r-1}(\sqrt{2} g \, t)}{\sqrt{2} g \, t}\, . \nonumber
\end{equation}
In particular, using this result for $r \, = \, 2$
we obtain the expression \eqref{allloopengscaling} for the
energy $E(g)$ in Fourier space.
As a further corrollary we find the Fourier transform of the third term
on the r.h.s.~of
\eqref{2loopintegral2}, as stated in \eqref{prefredholm}:
\begin{equation}
e^{-|t|/2} \, \int_{-\infty}^\infty du \, e^{- i \, t \, u} \; \,
\frac{1}{2} \, \partial_u \left(\frac{1}{x^+(u)} \, +
\frac{1}{x^-(u)} \right) \, = \, 2 \, \pi \, e^{-|t|} \, \frac{J_1(\sqrt{2} g
\, |t|)}{\sqrt{2} g}\, .
\end{equation}

\subsection{The String Dressing Factor}

In order to include the {\it dressing factor} for the
``string Bethe ansatz'' \cite{AFS,BS2}, as needed in the discussion
at the end of section \ref{weakcoupling}, we replace in
equation (\ref{2loopintegral2})
the kernel $K(u, \, u')$ from the gauge theory
Bethe ansatz by
\begin{equation}
K_s(u, \, u') \, = \, - \, \partial_u \, (u \, - \, u') \, \log
\left( \frac{ \bigl( 1 \, - \, g^2 \, / \, 2 \,
x^+(u) \, x^-(u') \bigr) \bigl( 1 \, - \, g^2 \, / \, 2 \, x^-(u) \,
x^+(u') \bigr) }{ \bigl( 1 \, - \, g^2 \, / \, 2 \, x^+(u) \,
x^+(u') \bigr) \bigl( 1 \, - \, g^2 \, / \, 2 \, x^-(u) \,
x^-(u') \bigr) } \right)^2 \, ,
\end{equation}
which will again be needed in a $u' \, \leftrightarrow \, - \, u'$ symmetrized
form.
In view of the analysis in the last section, the question arises as to whether
this expression can also be written as a one-parameter integral over pairs of
the form $K^\pm_j(u) \, K^\pm_j(u')$. As we shall see shortly, this is indeed
the case.

Quite clearly we have to deal with two distinct pieces, namely the part
involving $\partial_u \, u$ and that with $\partial_u \, u'$. In the first
case, the expression is explicitly symmetrized in $u'$ whereas $\partial_u \,
u$ on the whole is also even with respect to the integrals on the
half axis that we may expect to find. We are led to look for combinations
involving $K_j^+(u) \, + \, K_j^-(u)$ and likewise in $u'$. We remark that
under the Fourier transform $\partial_u \, u  \leftrightarrow  - \, t \,
\partial_t$. The differential operator can thus be incorporated at no expense.
Surprisingly, the $K_0^\pm$ alone suit our purpose: In the same fashion as
before we may demonstrate
\begin{eqnarray}
&& - \, \frac{1}{2} \, \partial_u \, u \, \log
\left( \frac{ \bigl( 1 \, - \, g^2 \, / \, 2 \,
x^+(u) \, x^-(u') \bigr) \bigl( 1 \, - \, g^2 \, / \, 2 \, x^-(u) \,
x^+(u') \bigr) }{ \bigl( 1 \, - \, g^2 \, / \, 2 \, x^+(u) \,
x^+(u') \bigr) \bigl( 1 \, - \, g^2 \, / \, 2 \, x^-(u) \,
x^-(u') \bigr) } \right)^2 \, + \, (u'  \leftrightarrow -u') \nonumber \\
& = & - \, g^2 \int_{-\infty}^\infty dt \, e^{i \, u \, t} \,
\int_{-\infty}^\infty dt' \, e^{i \, u' \, t'} \times \label{sp1} \\
&& \qquad 2 \, |t| \, \partial_{|t|} \, e^{- (|t| \, + \, |t'|)/2}
\int_0^1 \, d\lambda \, \lambda \, J_0(\sqrt{2} g
\lambda \, |t|)  \, J_0(\sqrt{2} g \lambda \, |t'|)  \nonumber \, .
\end{eqnarray}
For the second piece we must try $K_j^+(u) \, - \, K_j^-(u)$ and similarly
for $u'$, because the simple derivative in $u$ is odd while the extra power of
$u'$ forces antisymmetrization on the log factor. In a beautifully symmetric
way we can realize the term as a parameter integral this time over
antisymmetric combinations of only $K_1$:
\begin{eqnarray}
&& \frac{1}{2} \, \partial_u \, u' \, \log
\left( \frac{ \bigl( 1 \, - \, g^2 \, / \, 2 \,
x^+(u) \, x^-(u') \bigr) \bigl( 1 \, - \, g^2 \, / \, 2 \, x^-(u) \,
x^+(u') \bigr) }{ \bigl( 1 \, - \, g^2 \, / \, 2 \, x^+(u) \,
x^+(u') \bigr) \bigl( 1 \, - \, g^2 \, / \, 2 \, x^-(u) \,
x^-(u') \bigr) } \right)^2 \, + \, (u'  \leftrightarrow -u') \nonumber \\
& = & - \, g^2 \int_{-\infty}^\infty dt \, e^{i \, u \, t} \,
\int_{-\infty}^\infty dt' \, e^{i \, u' \, t'} \times \label{sp2} \\
&& \qquad 2 \, |t| \, \partial_{|t'|} \, e^{- (|t| \, + \, |t'|)/2}
\int_0^1 \, d\lambda \, \lambda \, J_1(\sqrt{2} g
\lambda \, |t|)  \, J_1(\sqrt{2} g \lambda \, |t'|)  \nonumber \, .
\end{eqnarray}
In the right hand sides of the last two formulas the derivatives can either
fall upon the exponential or on the Bessel functions. Accordingly, we
reproduce the Fourier transformed gauge theory kernel
$\hat K$
and an additional piece $\sqrt{2}\, g \, \tilde K$
, defined as
\begin{eqnarray}
\tilde K(t, \, t') & = & - \, 2 \, \int_0^1
\, d\lambda \, \lambda \, \Bigl[ \partial_t \, J_0(\lambda \, t)  \,
J_0(\lambda \, t') \, + \, \partial_{t'} \, J_1(\lambda \, t )  \,
J_1(\lambda \, t') \Bigr] \\
& = & \frac{t \, \bigl[ J_2(t) \, J_0(t') \, - \,
J_0(t) \, J_2(t') \bigr]} {(t \,-  \,t')(t \, + \, t')}\, . \nonumber
\end{eqnarray}
Here one should first do the parametric integration in both terms separately
and then differentiate and simplify.

Equation (\ref{finalform}) is replaced by:
\begin{eqnarray}
&& e^{-|t|/2} \, \int_{-\infty}^\infty du \, e^{- i \, t \, u} \;
\int_{-\infty}^{\infty} du' \, K_s(u, \, u') \; \sigma(u') \; = \\
&& 2 \, \pi \, g^2 \, |t| e^{-|t|} \int_{-\infty}^\infty dt' \,
\bigr[ \hat K(\sqrt{2} g |t|, \sqrt{2} g |t'|)
+ \sqrt{2} g \, \tilde K(\sqrt{2} g |t|, \sqrt{2} g |t'|) \,
\bigr] \; \hat \sigma(t') \, .
\nonumber
\end{eqnarray}



\end{document}